\documentclass[12pt]{article}

\usepackage{epsfig,amsmath,amssymb,graphics,graphicx} 

\usepackage{amscd}
\usepackage{amsfonts}

\usepackage{amsmath,amsfonts,amsthm,amssymb,eucal}
\usepackage{breqn}

\newtheorem{Definition}{Definition}[section]
\newtheorem{Theorem}{Theorem}[section]

\newtheorem{Corollary}{Corollary}[section]

\newtheorem{Remark}{Remark}[section]

\begin{document}

\begin{center}

Communications in Nonlinear Science and Numerical Simulation 102 (2021) 105947

\vskip 7mm

{\bf \large Fractional Dynamics with Non-Local Scaling}
\vskip 3mm

\vskip 7mm
{\bf \large Vasily E. Tarasov}$^{1,2}$ \\
\vskip 3mm

${}^1$ {\it Skobeltsyn Institute of Nuclear Physics, \\ 
Lomonosov Moscow State University, Moscow 119991, Russia} \\
{E-mail: tarasov@theory.sinp.msu.ru} \\

${}^2$ {\it Faculty "Information Technologies and Applied Mathematics", \\ Moscow Aviation Institute (National Research University), Moscow 125993, Russia } \\

\begin{abstract}
Nonlinear fractional dynamics with scale invariance in continuous and discrete time approaches are described.
We use non-integer-order integro-differential operators that can be interpreted as generalizations of scaling (dilation) differential operator for the case of non-locality.
Nonlinear integro-differential equations with Hadamard type operators of non-integer orders with respect to time and periodic sequence of kicks are considered.
Exact solutions of these equations are derived without using approximations.
Using these solutions for discrete time points, we derive
mappings with non-local scaling in time from proposed equations without approximation.
Non-local mappings are obtained in general for arbitrary orders of the Hadamard type fractional operators. 
An example of these mappings with non-local scaling in time is given for arbitrary positive orders of integro-differential equations with kicks. 
The proposed mappings are independent of the period of kicks at zero initial conditions.
The proposed approach can be used to describe the nonlinear fractional dynamics that is characterized by scale invariance. 
\end{abstract}

\end{center}

\noindent

\noindent


\section{Introduction}

Scaling properties can be described by the transformation of a function $f(x)$ under the change of the variable $x$ by some scale factor $\rho$: $x \to \rho \, x$.
Scaling and scale invariance are important for various applications. 
For example, in modern physics, it is used the group $C(1,3)$ of conformal transformations of the space-time $\mathbb{R}^4_{1,3}$. These transformations, which describe the symmetries of the Maxwell's equations of of electromagnetic fields in vacuum \cite{EPJPlus2020}, contain the inhomogeneous Lorentz transformations,
the special conformal transformations and
the dilatations (scaling) \cite[p.409]{Barut}. 
The dilatations (scaling) $\Pi_{\rho}$ is defined as
\begin{equation} 
\Pi_{\rho} \, f(x^{\mu}) = f(\rho \, x^{\mu}) .
\end{equation}
For one-dimensional space $\mathbb{R}^1$ the dilation \cite[p.95]{FC1} is
\begin{equation}
\Pi_{\rho} \, f(t) = f(\rho \, t) ,
\end{equation}
where $t\in \mathbb{R}$ and $\rho>0$.
For the conformal group $C(1,3)$ on the space of scalar functions over the space-time $\mathbb{R}^4_{1,3}$, basis elements of the Lie algebra are represented by the differential operators: 
the generators of the Poincare subalgebra $M_{\mu,\nu}$, $P_{\mu}$, 
the generator $K_{\eta}$ of the special conformal transformations and the differential operator 
\begin{equation}
D = \sum^{3}_{\mu=0} x^{\mu} \frac{\partial}{\partial x^{\mu}}
\end{equation}
that generates the dilatations (dilations, scaling) \cite[p.412]{Barut}. 
For one-dimensional space $\mathbb{R}^1$, the differential operator of dilation (scaling) \cite[p.95]{FC1} is
\begin{equation}
{\cal S}_t = t \, \frac{d}{dt} .
\end{equation}
Usually, the differential operators of scaling (dilation) are described as operators of integer orders ${\cal S}^m_t=({\cal S}_t)^m$, where $m \in \mathbb{N}$.
Obviously, the operators are scale invariant
\begin{equation}
\Pi_{\rho} \, {\cal S}^m_t \, f(t) = 
{\cal S}^m_t \, \Pi_{\rho} \, f(t) .
\end{equation}
As another example of a scalar invariant differential operator, we indicate the operator 
\begin{equation} \label{HT-Dm}
{\cal D}^{t,m}_{a,\eta} = 
t^{-\eta} \, {\cal S}^m_t \, t^{\eta}: \quad
\Pi_{\rho} \, {\cal D}^{t,m}_{a,\eta} \, f(t) = 
{\cal D}^{t,m}_{a,\eta} \, \Pi_{\rho} \, f(t) ,
\end{equation}
where $\rho>0$ and $t>a>0$.
The operator ${\cal D}^{t,m}_{a,\eta}$ is local, since the action of this operator on a function is determined by the properties of the function in an infinitesimal neighborhood of the point under consideration. 


In fractional dynamics \cite{Springer2010}, the differential operators of non-integer orders 
\cite{FC1,FC2,FC3,FC4,FC5,Handbook2019-1,Handbook2019-2} are used to describe non-locality in space and time \cite{Handbook2019-4,Handbook2019-5}, \cite{BOOK-MDPI-2020,BOOK-DG-2021}.
For some integro-differential operators of non-integer order, 
the scaling properties are described. 
For example, the Riemann-Liouville integration of order $\alpha$ is characterized \cite[p.96]{FC1} by the following scaling property 
\begin{equation} \label{Pi-INF}
\Pi_{\rho} \, I^{t,\alpha}_{RL,0+} [\tau] \, f(\tau) = 
\rho^{\alpha} \, I^{\alpha}_{RL,0+} [\tau] \, \Pi_{\rho} \, f(\tau) .
\end{equation}
For the Marchaud fractional derivative, the scaling property
\begin{equation} 
\Pi_{\rho} \, D^{\alpha}_{M,0+} [\tau] \, f(\tau) = 
\rho^{-\alpha} \, D^{\alpha}_{M,0+} [\tau] \, \Pi_{\rho} \, f(\tau) 
\end{equation}
holds for $\alpha>0$ (see equation 5.62 in \cite[p.111]{FC1}).

Because of this, the question arises about the generalization of the differential operator of dilation 
${\cal S}^m_t=({\cal S}_t)^m$ from 
integer order $m \in \mathbb{N}$ to arbitrary positive orders
$\alpha \in \mathbb{R}_{+}$ to describe dynamics with non-local scaling in time and space.
 

We can consider a generalization of the differential operator of scaling (dilation) for $\mathbb{R}$ in the form 
\begin{equation} \label{delta-K}
{\cal S}^{t,m}_{(K)} [\tau] \, f(\tau) = \int^t_{a} d\tau \, K(t,\tau) \, ({\cal S}_{\tau})^{m} \, f(\tau) ,
\end{equation}
where $m \in \mathbb{N}$ and $K(t,\tau)$ is the kernel that characterizes the non-locality.
To describe fractional dynamics with non-local scaling, 
we should impose the condition 
\begin{equation}
\Pi_{\rho} \, {\cal S}^{t,m}_{(K)}[\tau]  \, f(\tau) = 
{\cal S}^{t,m}_{(K)}[\tau]  \, \Pi_{\rho} \, f(\tau) .
\end{equation}
In order to have a calculus for the integro-differential operators of non-local scaling, we should define the
integral operator
\begin{equation} \label{INT-K}
{\cal J}^{t}_{(K)}[\tau] \, f(\tau) = 
\int^t_{a} d\tau \, K_J(t,\tau) \, f(\tau) ,
\end{equation}
such that
\begin{equation}
\Pi_{\rho} \, {\cal J}^{t}_{(K)} [\tau] \, f(\tau) = 
{\cal J}^{t}_{(K)}[\tau] \, \Pi_{\rho} \, f(\tau) ,
\end{equation}
and
\begin{equation}
{\cal S}^{t,n}_{(K)}[\tau] \, {\cal J}^{\tau}_{(K)}[s] \, f(s) = f (t) .
\end{equation}

The general solution of such a problem, that is, the problem of finding types of the operator kernels $K(t,\tau)$ and $K_J(t,\tau)$ to satisfy these conditions, has not been solved in the general case at present time. 
However, there are examples of fractional integro-differential operators that satisfy these conditions and form a fractional calculus \cite{FC1,FC4}. 
These operators are called the Hadamard fractional integrals and derivatives (see Section 18.3 in \cite[pp.329-333]{FC1} and Section 2.7 in \cite[pp.110-120]{FC4}).
The Hadamard fractional calculus is actively developing at present time in mathematics, 
including generalization to the Hadamard-type fractional calculus 
\cite{BKT2002-1,Kilbas2001,KilbasTitjura2002,BKT2002-2,BKT2002-3,KMT2003,KMT2003E},
its Caputo modifications \cite{Jarad2012,Jarad2014,Zafar2020},
and other generalizations \cite{ChangpinLi,Garra2013,Garra2018}.
 
For example, if we use the kernel
\begin{equation}
K(t,\tau) = \frac{1}{\Gamma (m-\alpha)} \frac{1}{\tau}
\left( \ln \left( \frac{\tau}{t} \right) \right)^{m-\alpha+1} ,
\end{equation}
where $m-1 < \alpha < m$, then expression \eqref{delta-K} defines 
the Hadamard fractional derivative (HFD) of the order $\alpha \in \mathbb{R}_{+}$.

For the kernel 
\begin{equation}
K_J(t,\tau) = \frac{1}{\Gamma (\alpha)} \frac{1}{\tau}
\left( \ln \left( \frac{\tau}{t} \right) \right)^{\alpha-1} ,
\end{equation}
where $\alpha>0$, equation \eqref{INT-K} defines
the Hadamard fractional integrals (HFI) of the order $\alpha>0$ in the form
\begin{equation} \label{DEF-INT-H}
{\cal J}^{t,\alpha}_{a,0} [\tau] f (\tau) =
\frac{1}{\Gamma(\alpha)} \int^{t}_{a} \frac{d\tau}{\tau} 
\left(\ln \frac{t}{\tau}\right)^{\alpha-1} \, f(\tau) \, ,
\end{equation}
where $t \in (a,b)$ with $0 \le a < b \le \infty$.

For the HFI, the following scaling property
\begin{equation} 
\Pi_{\rho} {\cal J}^{t,\alpha}_{a,0}[\tau] f(\tau) = 
{\cal J}^{t,\alpha}_{a,0} [\tau] \Pi_{\rho} f(\tau) 
\end{equation}
is satisfied (see equation 5.62 in \cite[p.111]{FC1}).
In addition, the HFI satisfies the property
\begin{equation}
{\cal S}_t \, {\cal J}^{t,\alpha+1}_{a,0}[\tau] f(\tau) =
t \frac{d}{dt} {\cal J}^{t,\alpha+1}_{a,0}[\tau] f(\tau) =
{\cal J}^{t,\alpha}_{a,0} f(t) 
\end{equation}
for $\alpha>0$ (see equation 18.53 in \cite[p.331]{FC1}).

The HFD is defined (see Eq. 18.54 in \cite[p.332]{FC1}) by the equation
\begin{equation}
{\cal D}^{t,\alpha}_{a,0}[\tau] f(\tau) = 
{\cal J}^{t,1-\{\alpha\}}_{a,0} [\tau]
\left( \tau \frac{d}{d\tau} \right)^{[\alpha]+1} f(\tau) =
\left( t \frac{d}{dt} \right)^{[\alpha]+1} 
{\cal J}^{t,1-\{\alpha\}}_{a,0}[\tau] f(\tau) 
\end{equation}
for $\alpha>0$, where $[\alpha]$ is the integer part of $\alpha$, and $\{\alpha\}=\alpha-[\alpha]$.
If the dilatation operator of integer order ${\cal S}^{[\alpha]+1}$ acts first on a function $f$, then the integro-differential operator is often called the Caputo modification 
\cite{Jarad2012,Jarad2014}.
 
Note that fractional operators with continuously distributed scaling (dilation) $\Pi_{\rho}$, where $\rho$ is a random variable, are suggeted in Section 9 of \cite{CNSNS-2019-LAG} and \cite{PFDA-2019} as generalizations of the Erdelyi-Kober fractional operators. 


In this article, to consider fractional dynamics with non-local scaling (dilation) in the discrete and continuous time approach, we will use a generalization of the Hadamard fractional operators and integer-order operator \eqref{HT-Dm} that is proposed in \cite{BKT2002-1,Kilbas2001}. 

To derive the discrete-time fractional dynamics with non-local scaling, we will use the approach that was first proposed in 
\cite{Tarasov-Zaslavsky,Tarasov-Map1,Tarasov-Map2} 
(see also \cite{Springer2010} and \cite{Tarasov-Map3,TT-Logistic,Entropy2021,Mathematics2021,MMAS2021,Mathematics2021-2}). 
This approach allows us to derive discrete-time mappings with non-locality in time from integro-differential equations of non-integer orders without approximation. 
To derive mapping with non-local scaling, we use exact solutions of nonlinear non-integer-order differential equations with kicks.
The proposed mappings describe the fractional dynamics with non-local scaling in discrete time approach.

In continuous-time fractional dynamics, the non-locality in time means that the behavior of the system depends on the history of changes in this system in the past over a finite time interval. 
In this case, the system should be described by integro-differential equations that cannot be represented in the form of finite number of differential equations of integer orders.
The scaling property of continuous-time approach is described by the integro-differential operators, which are generalization of the dilation (scaling) differential operator of integer order, which form a calculus.

The locality of discrete-time mappings means that the next step $X_{n+1}$ is defined only by the previous step $X_{n}$ (or a fixed number of previous steps, which represent time delays) , i.e. 
$X_{n+1}={\cal F}(X_n)$ (or $X_{n+1}={\cal F}(X_n,X_{n-1})$ for example). 
In discrete-time fractional dynamics, the non-locality in time means that the next step $X_{n+1}$ depends on all previous steps, i.e. 
$X_{n+1} = {\cal F} (X_n,X_{n-1}, ...,X_1)$,
and the number of variables in the function ${\cal F}$ increases with each new step. 

The discrete-time non-local mappings, which are derived from equations with integro-differential operators and periodic kicks, are independent of the kick period at least at zero initial conditions. 

To describe the fractional dynamics with continuous time, we will use non-local generalizations of 
the differential equation of the second order with periodic kicks
\begin{equation} \label{Eq-Integer}
\frac{d^2 X(t)}{dt^2}+ A \, \Lambda[X(t)] \sum^{\infty}_{k=1} \delta \Bigl(\frac{t}{\theta}-k \Bigr)=0 ,
\end{equation}
where $\theta$ is the period of kicks, $A$ is the kick amplitude, and $\Lambda [X]$ is a real-valued function.
It is known (Sec. 5 in \cite{SUZ},
Sec. 5.2, 5.3 in \cite[pp.60-68]{Zaslavsky2}, and Sec. 1.2 in \cite[pp.16-17]{Schuster}, Ch. 18 in \cite[pp.409-453]{Springer2010}) that equation \eqref{Eq-Integer} can be represented by the discrete-time dynamical mapping 
\begin{equation} \label{Map-Integer}
X_{n+1} = X_n + X^{(1)}(0) \, \theta - 
A \, \theta^2 \, \Lambda [X_n] , 
\end{equation}
where
\begin{equation} 
X_n = \lim_{\epsilon \rightarrow 0+} X(n \theta -\epsilon) .
\end{equation}
Equation of motion \eqref{Eq-Integer} is usually considered in the Hamiltonian form with the momentum $Y(t)=X^{(1)}(t)$.
In this case, the mapping is presented by two-dimenisional mappings \cite{Zaslavsky2,SUZ,Springer2010}, which are given 
by equations \eqref{Map-Integer} and 
$Y_{n+1}=Y_n- A \, \theta \, \Lambda [X_n]$, 
\cite[pp.411-412]{Springer2010}.
In equation \eqref{Map-Integer}, we see that this mapping is local and depends on the kick period.
Equation \eqref{Eq-Integer} and corresponding maps \eqref{Map-Integer} are actively used in chaos theory and nonlinear dynamics for modeling various physical systems 
\cite{SUZ,Zaslavsky2,Chirikov,Schuster,Springer2010}. 

In this work, we obtain the equations of fractional dynamics with discrete time in the form of non-local mappings, which do not depend on the kick period $\theta$.
These mappings describe solution of non-local generalizations of equation \eqref{Eq-Integer} without approximation. 
This goal is achieved by using the approach suggested in 
\cite{Tarasov-Zaslavsky,Tarasov-Map1,Tarasov-Map2,Springer2010}.


\section{Hadamard type fractional operators and their properties}

In this section, we present the definitions and some properties of the Hadamard type fractional integral (HTFI) and fractional differential (HTFD) operators that are needed in the next sections. 
These operators were first suggested by Paul L. Butzer, Anatoly A. Kilbas and Juan J. Trujillo in 2002 \cite{BKT2002-1,Kilbas2001}, and therefore they can be called the Butzer-Kilbas-Trujillo operators. 
The properties of the HTFD and HTFI operators are described in 
papers \cite{BKT2002-1,Kilbas2001,KilbasTitjura2002,BKT2002-2,BKT2002-3,KMT2003,KMT2003E}.

\subsection{Hadamard type fractional integrals}

The HTFI operator of the order $\alpha>0$ is given by the following definition \cite{BKT2002-1,Kilbas2001}.

\begin{Definition}
The Hadamard type fractional integral (HTFI) operator of the order $\alpha>0$ is defined by the expression
\begin{equation} \label{DEF-INT-HT}
{\cal J}^{t,\alpha}_{a,\eta}[\tau] X(\tau) = 
\frac{1}{\Gamma(\alpha)} \int^{t}_{a} \frac{d\tau}{\tau} \, 
\left( \frac{\tau}{t}\right)^{\eta}
\left( \ln \frac{t}{\tau}\right)^{\alpha-1} \, X(\tau) ,
\end{equation}
where $\eta \in \mathbb{R}$, $t \in (a,b)$, $a>0$.
\end{Definition}

\begin{Remark}
The operator ${\cal J}^{t,\alpha}_{a,\eta}$ is bounded in the space $X^p_c(a,b)$, where $\eta \in \mathbb{R}$, $\eta >c$, $p\ge 1$, $a>0$ (see Theorem 2.1 in \cite[p.1194]{Kilbas2001}).
The space $X^p_c(a,b)$ with $c \in \mathbb{R}$ and $p \ge 1$ 
is the weighted $L_p$-space with the power weight, which consists of those complex-valued Lebesgue measurable functions $X(t)$ on $(a,b)$ for which 
\begin{equation} 
||X||_{X^p_c} = 
\left( \int^b_a \frac{d\tau}{\tau} |\tau^c X(\tau)|^p \right)^{1/p} < \infty .
\end{equation} 
In particular, when $c=1/p$, the space $X^p_c(a,b)$ 
coincides with the space $L_p(a,b)$, i.e., 
$X^p_{1/p}(a,b) =L_p(a,b)$.
\end{Remark}

\begin{Remark}
Operators \eqref{DEF-INT-HT} satisfy the following scaling property
\begin{equation} 
\Pi_{\rho} \, {\cal J}^{t,\alpha}_{a,\eta}[\tau] X(\tau) = 
{\cal J}^{t,\alpha}_{a,\eta}[\tau] \, \Pi_{\rho} \, X(\tau) ,
\end{equation}
where $\rho>0$, $\alpha>0$, $a>0$, $t \in (a,b)$.
\end{Remark}

For $\alpha = m \in \mathbb{N}$, operator \eqref{DEF-INT-HT} has \cite[p.~1191]{Kilbas2001} the form
\[
{\cal J}^{t,m}_{a, \eta} [\tau] X(\tau) =
t^{-\eta} \int^{t}_{a} \frac{d t_1}{t_1} 
\int^{t_1}_{a} \frac{d t_2}{t_2} ...
\int^{t_{m-1}}_{a} t^{\eta}_m X(t_m) \frac{d t_m}{t_m} =
\]
\begin{equation} \label{DEF-INT-HT-m}
=\frac{1}{(m-1)!} 
\int^{t}_{a} \left( \frac{\tau}{t}\right)^{\eta}
\left( \ln \frac{t}{\tau}\right)^{m-1} X(\tau) \, \frac{d \tau}{\tau} .
\end{equation}

\begin{Remark}
Integral operator \eqref{DEF-INT-HT} with $\eta=0$ is called the Hadamard FI, which was proposed by Jacques S. Hadamard \cite{Hadamard1892} in 1892. 
These operators and their properties are described in \cite[pp.~110-120]{FC4} (see also Sections 18.3, 23.1 of \cite{FC1} and \cite{AANT2017}).
\end{Remark}


\subsection{Hadamard type fractional differential operator}

Let us give the definition of the HTFD operator (for example, see Section 18.3 in \cite{FC1} and \cite[pp.~11-112]{FC4}). 

\begin{Definition}
The Hadamard type fractional differential (HTFD) operator of the order $\alpha \in (m-1,m)$, $m \in \mathbb{N}$ is defined by the expression
\begin{equation} \label{DEF-DER-HT}
{\cal D}^{t,\alpha}_{a,\eta}[\tau] X(\tau) = 
t^{-\eta} \left(t \frac{d}{dt}\right)^m t^{\eta} 
{\cal J}^{t,m-\alpha}_{a,\eta} [\tau] X(\tau) =
\end{equation}
\[
\frac{1}{\Gamma(m-\alpha)} t^{-\eta} 
\left(t \frac{d}{dt}\right)^m t^{\eta}
\int^{t}_{a} \frac{d\tau}{\tau} 
\left( \frac{\tau}{t}\right)^{\eta}
\left( \ln \frac{t}{\tau}\right)^{m-\alpha+1} \, X(\tau) ,
\]
where ${\cal J}^{t,m-\alpha}_{a,\eta}$ is the HTFI operator, $t>a>0$.
\end{Definition} 

For $\alpha=m \in \mathbb{N}$, the HTFD operator is the integer-order differential operator \cite[p.~112]{FC4} in the form
\begin{equation} \label{EQ-HT-m}
{\cal D}^{t,m}_{a,\eta}[\tau] X (\tau) = 
t^{-\eta} \left(t \frac{d}{dt}\right)^m \left( t^{\eta} \, X(t) \right) .
\end{equation}
Since operator \eqref{EQ-HT-m} is local, we will use the notation
\[
{\cal D}^{t,m}_{a,\eta} X(t) =
{\cal D}^{t,m}_{a,\eta}[\tau] X (\tau) ,
\]
if $m \in \mathbb{N}$.

\begin{Remark}
The operator ${\cal D}^{t,\alpha}_{a,\eta}$ exists almost everywhere on the space $AC^m_{{\cal S},\eta}(a,b)$. 
This statement is proved as Theorem 3.2 in \cite[p.1198]{Kilbas2001}.
The space $AC^{m}_{{\cal S}, \eta}[a,b]$ consists of functions $X(t)$ on $[a,b]$ that have ${\cal S}^{k}_t (t^{\eta} \, X(t))$
for $k=1,...m-1$, and
${\cal S}^{m-1}_t (t^{\eta} \, X(t))$ is absolutely continuous on $[a,b]$, where ${\cal S}$ denotes the operator ${\cal S}_t = t \, d/dt$ (see \cite[p.1193]{Kilbas2001}).
\end{Remark}

\begin{Remark}
Operator \eqref{DEF-DER-HT} with $\eta=0$ is called the Hadamard fractional derivatives\cite[pp.~110-120]{FC4}, \cite{AANT2017}.
\end{Remark}

\begin{Remark}
For $\eta=0$, and $\alpha=m \in \mathbb{N}$, the HTFD operator is the differential operator of scaling (dilation) of the integer order 
\begin{equation} 
{\cal D}^{t,m}_{a,0} X (t) = {\cal S}^m_t \, X(t) =
\left(t \frac{d}{dt}\right)^m \, X(t) .
\end{equation}
\end{Remark}

\begin{Definition}
The Caputo modification of the HTFD operator of the order $m-1<\alpha <m$ is defined as
\begin{equation} \label{DEF-DER-HT-C}
^C{\cal D}^{t,\alpha}_{a,\eta} [\tau] X(\tau) = 
{\cal J}^{t,m-\alpha}_{a,\eta} [\tau]
\left( \tau^{-\eta} \left(t \frac{d}{d\tau}\right)^m \tau^{\eta} \right)
X (\tau) =
\end{equation}
\[
\frac{1}{\Gamma(m-\alpha)} \int^{t}_{a} \frac{d\tau}{\tau} 
\left( \frac{\tau}{t}\right)^{\eta}
\left( \ln \frac{t}{\tau}\right)^{m-\alpha+1} \, 
\left( \tau^{-\eta} \left(\tau \frac{d}{d\tau}\right)^m \tau^{\eta} \right)
X(\tau) ,
\]
where ${\cal J}^{t,m-\alpha}_{a,\eta}$ is the HTFI operator, $t>a>0$.
\end{Definition} 

\begin{Remark}
Operators \eqref{DEF-DER-HT} and \eqref{DEF-DER-HT-C} have the following scaling property
\begin{equation} 
\Pi_{\rho} \, {\cal D}^{t,\alpha}_{a,\eta}[\tau] X(\tau) = 
{\cal D}^{t,\alpha}_{a,\eta}[\tau] \, \Pi_{\rho} \, X(\tau) ,
\end{equation}
\begin{equation} 
\Pi_{\rho} \, ^C{\cal D}^{t,\alpha}_{a,\eta}[\tau] X(\tau) = 
\, ^C{\cal D}^{t,\alpha}_{a,\eta}[\tau] \, \Pi_{\rho} \, X(\tau) ,
\end{equation}
where $\rho>0$, $\alpha>0$, $a>0$, $t \in (a,b)$.
\end{Remark}


\subsection{Equivalence of integro-differential equations}

The HTFD and HTFI operators form a fractional calculus (FC) \cite{BKT2002-1,Kilbas2001,BKT2002-1,Kilbas2001,KilbasTitjura2002,BKT2002-2,BKT2002-3,KMT2003,KMT2003E}.
The fundamental theorems of FC for the HTFD and HTFI operators are described by Lemma 3 and Lemma 5 in \cite[p.~735]{KMT2003} (see also Theorem 4.7 of \cite[p.~1203]{Kilbas2001}, and \cite{KilbasTitjura2002,KMT2003E}).

Anatoly A. Kilbas, Sergei A. Marzan, and Aleksandr A. Tityura, 
\cite{KMT2003,KMT2003E} investigated the Cauchy problem for the equations with the HTFD operators. 
Theorem 1 in \cite[p.~736]{KMT2003}) gives the equivalence of the Cauchy problem for equation with HTFD operator and the Volterra integral equation of the 2nd kind. 

\begin{Theorem} \label{Theorem-3-1}
Let $X(t) \in AC^{m}_{{\cal S},\eta}(a,b)$, and 
$F[t,X(t)] \in X^p_{c}(a,b)$ with $c>\eta$, $p\ge 1$, 
$m-1<\alpha \le m$, and $a>0$. 
Then the Cauchy problem for the integro-differential equation
\begin{equation} \label{Cauchy-HT-1}
{\cal D}^{t,\alpha}_{a,\eta}[\tau] X(\tau) = F[t,X(t)] ,
\end{equation}
and the conditions
\begin{equation} \label{Cauchy-HT-2}
\lim_{t\to a+} t^{\eta} {\cal D}^{t,m-k}_{a,\eta} 
{\cal J}^{t,m-\alpha}_{a,\eta}[\tau] X(\tau) = C_k , 
\quad (k=1, ..., m) ,
\end{equation}
where $C_k \in \mathbb{R}$, is equivalent to the Volterra integral equation of the second kind
\begin{equation} \label{Cauchy-HT-3}
X(t) = \sum^{m}_{k=1} \frac{C_k \, t^{-\eta}}{\Gamma(\alpha-k+1)} 
\left( \ln \frac{t}{a}\right)^{\alpha-k}
+ \frac{1}{\Gamma(\alpha)} \int^t_a \frac{d\tau}{\tau} 
\left( \frac{\tau}{t}\right)^{\eta} 
\left( \ln \frac{t}{\tau}\right)^{\alpha-1} \, F[\tau,X(\tau)] .
\end{equation}
\end{Theorem}

\textbf{Proof.}
This theorem was proved in paper \cite{KMT2003} (see Theorem 1 in \cite[p.~736]{KMT2003}).

\ \ \ $\Box$ \\

\begin{Remark}
Note that conditions \eqref{Cauchy-HT-2} with $k=1, ..., m-1$ 
can be written as
\begin{equation} \label{Cauchy-HT-2b}
\lim_{t\to a+} 
 t^{\eta}{\cal D}^{t,\alpha-k}_{a,\eta}[\tau] X(\tau) = C_k , 
\end{equation}
and for $k=m$ as
\begin{equation} \label{Cauchy-HT-2c}
\lim_{t\to a+} t^{\eta}{\cal J}^{t,m-\alpha}_{a,\eta} [\tau] X(\tau) = C_m . 
\end{equation}
\end{Remark}

Equivalence, which is described by Theorem \ref{Theorem-3-1}, means that function $X(t)$ is a solution to the Cauchy problem \eqref{Cauchy-HT-1}, \eqref{Cauchy-HT-2} almost everywhere, if $X(t)$ is a solution to the Volterra integral equation \eqref{Cauchy-HT-3}. 

Note that equation \eqref{Cauchy-HT-3} can be written by using HTFI operator \eqref{DEF-INT-HT} in the form 
\begin{equation} \label{Cauchy-HT-3b}
X(t) = t^{-\eta}\sum^{m}_{k=1} \frac{C_k}{\Gamma(\alpha-k+1)} 
\left( \ln \frac{t}{a}\right)^{\alpha-k}
+ {\cal J}^{t,\alpha}_{a,\eta}[\tau] F[\tau,X(\tau)] .
\end{equation}

\begin{Theorem} \label{Theorem-3-2}
Let $X(t) \in AC^{m}_{{\cal S},\eta}(a,b)$, and $F[t,X(t)] \in X^p_{c}(a,b)$ with $c>\eta$, $p\ge 1$, $m-1<\alpha \le m$, and $a>0$. 
Then the Cauchy problem for the equation
\begin{equation} \label{Cauchy-HT-1C}
^C{\cal D}^{t,\alpha}_{a,\eta}[\tau] X(\tau) = F[t,X(t)] ,
\end{equation}
and the conditions
\begin{equation} \label{Cauchy-HT-2C}
\lim_{t\to a+} D^{t,k}_{a,\eta} X(t) = C_k^{*} , \quad (k=0, ..., m-1) 
\end{equation}
is equivalent to the Volterra integral equation of the second kind in the form
\begin{equation} \label{Cauchy-HT-3C}
X(t) = \sum^{m-1}_{k=0} \frac{C_k^{*}}{\Gamma(k+1)} 
\left( \frac{a}{t}\right)^{\eta}
\left( \ln \frac{t}{a}\right)^{k} + 
\frac{1}{\Gamma(\alpha)} \int^t_a \frac{d\tau}{\tau} 
\left( \frac{\tau}{t}\right)^{\eta} 
\left( \ln \frac{t}{\tau}\right)^{\alpha-1} \, F[\tau,X(\tau)] .
\end{equation}
\end{Theorem}

\textbf{Proof.}
The proof of this theorem follows from the second fundamental theorem of the FC that is given in Theorem 4.3 of \cite[p.8]{Zafar2020}.

\ \ \ $\Box$ \\


\section{Fractional dynamics: from continuous to discrete time}

Let us consider the equation with HTFD operator
\begin{equation} \label{FDE-HT1}
{\cal D}^{t,\alpha}_{a,\eta}[\tau] X(\tau) +
A \, \Lambda [X(t-\epsilon)] \sum^{\infty}_{k=1} \delta \Bigl(\frac{t}{\theta}-k \Bigr)=0 ,
\end{equation}
where ${\cal D}^{t,\alpha}_{a,\eta}$ is operator \eqref{DEF-DER-HT} of the order $\alpha \in (m-1,m]$, and $0<a<\theta$. 
Here $\theta$ is period of the periodic sequence of kicks, 
$A$ is an amplitude of the kicks, 
$\Lambda [X]$ is some real-valued function and 
$\Lambda [X(t-\epsilon)] \in C^{\infty}(a,b)$.

Equation \eqref{FDE-HT1}, which contains the delta functions that are distributions (the generalized functions), 
should be considered in a generalized sense (see Sec 8 \cite[pp.145-160]{FC1} and \cite{GenFunc1,GenFunc2}), 
i.e. on the test function spaces.
We also use $t-\epsilon$ instead of $t$, to make a sense of the 
product of the delta function and $\Lambda [X]$ for the case $X(t_k-0) \neq X(t_k+0)$, where $t_k=k\theta$ \cite{Edelman5}.

\begin{Theorem} \label{Theorem-Xt-1}
Let $X(t) \in AC^{m}_{{\cal S},\eta}(a,b)$ 
with $m-1<\alpha \le m$, $a>0$,
and $\Lambda [X(t-\epsilon)] \in C^{\infty}(a,b)$.
Equation \eqref{FDE-HT1} 
with $a \in (0,\theta)$ and $\alpha \in (m-1,m]$
has the solution for $ t \in (n\theta,(n+1)\theta)$ in the form
\begin{equation} \label{FDE-HT4b}
X(t) = \sum^{m}_{j=1} \frac{C_j \, t^{-\eta}}{\Gamma(\alpha-j+1)} 
\left( \ln \frac{t}{a}\right)^{\alpha-j}
- \frac{A}{\Gamma(\alpha)} 
\sum^{n}_{k=1} \frac{1}{k} 
\left( \frac{k\, \theta}{t}\right)^{\eta} 
\left( \ln \frac{t}{k\, \theta}\right)^{\alpha-1} \Lambda [X(t_k-\epsilon)],
\end{equation}
where $t_k=k\theta$ and
\begin{equation} \label{FTFC-HT-3b}
C_j = \lim_{t \to a+} 
\left(t \frac{d}{dt}\right)^{m-j} t^{\eta}
{\cal J}^{t,m-\alpha}_{a,\eta} [\tau] X(\tau) , 
\quad (j=1,...,m) .
\end{equation}
\end{Theorem}

{\bf Proof.}
Equation \eqref{FDE-HT1} has the form of equation \eqref{Cauchy-HT-1} with the function
\begin{equation}
F[t,X(t)] = - A \, \Lambda [X(t-\epsilon)] \sum^{\infty}_{k=1} \delta \Bigl(\frac{t}{\theta}-k \Bigr) .
\end{equation}
Using Theorem \ref{Theorem-3-1} Cauchy problem \eqref{FDE-HT1}, \eqref{Cauchy-HT-3} is equivalent to the integral equation
\[
X(t) = \sum^{m}_{j=1} \frac{C_j \, t^{-\eta}}{\Gamma(\alpha-j+1)} 
\left( \ln \frac{t}{a}\right)^{\alpha-j} -
\]
\begin{equation} \label{FDE-HT2}
\frac{A}{\Gamma(\alpha)} \int^t_a \frac{d\tau}{\tau} 
\left( \frac{\tau}{t}\right)^{\eta} 
\left( \ln \frac{t}{\tau}\right)^{\alpha-1} 
\Lambda [X(\tau-\epsilon)] \sum^{\infty}_{k=1} \delta \Bigl(\frac{\tau}{\theta}-k \Bigr) ,
\end{equation}
where $m-1 < \alpha \le m$, $m \in \mathbb{N}$ and $C_j$ is defined by equation \eqref{Cauchy-HT-2}.
The equivalence of equations \eqref{FDE-HT1} and \eqref{FDE-HT2} should be considered in the generalized sense (on the space of test functions). 	 

For $n\theta<t<(n+1)\theta$, equation \eqref{FDE-HT2} becomes
\[
X(t) = \sum^{m}_{j=1} \frac{C_j \, t^{-\eta}}{\Gamma(\alpha-j+1)} 
\left( \ln \frac{t}{a}\right)^{\alpha-j} -
\]
\begin{equation} \label{FDE-HT3}
\frac{A}{\Gamma(\alpha)} \sum^{n}_{k=1} \int^t_a
\frac{d\tau}{\tau} \left( \frac{\tau}{t}\right)^{\eta} 
\left( \ln \frac{t}{\tau}\right)^{\alpha-1} 
\Lambda [X(\tau-\epsilon)] \delta \Bigl(\frac{\tau}{\theta}-k \Bigr) .
\end{equation}
Then we apply the property 
\begin{equation} \label{Eq-Proof-1c}
\int^t_0 d\tau \, g(\tau) \, \delta \Bigl(\frac{\tau}{\theta}- k \Bigr) = \theta \, g (\theta\ , k) ,
\end{equation}
which is satisfied if $f(\tau)$ is continuous function at $\tau=\theta \, k$ and $a< \theta \, k <t$ with $t_k=k\theta$.
For $n\theta<t<(n+1)\theta$ and $0<a<\theta$, equation \eqref{FDE-HT3} gives 
\begin{equation} \label{FDE-HT4}
X(t) = \sum^{m}_{j=1} \frac{C_j \, t^{-\eta}}{\Gamma(\alpha-j+1)} 
\left( \ln \frac{t}{a}\right)^{\alpha-j}
- \frac{A \, \theta}{\Gamma(\alpha)} 
\sum^{n}_{k=1} \frac{1}{k\, \theta} 
\left( \frac{k\, \theta}{t}\right)^{\eta} 
\left( \ln \frac{t}{k\, \theta}\right)^{\alpha-1} 
\Lambda [X(t_k-\epsilon)] .
\end{equation}

\ \ \ $\Box$ \\

\begin{Corollary}
For $\eta=0$, equation \eqref{FDE-HT1} is integro-differential equation with Hadamard derivative, for which the solution is given as
\begin{equation} \label{FDE-HT4c}
X(t) = \sum^{m}_{j=1} \frac{C_j}{\Gamma(\alpha-j+1)} 
\left( \ln \frac{t}{a}\right)^{\alpha-j}
- \frac{A}{\Gamma(\alpha)} 
\sum^{n}_{k=1} \frac{1}{k} 
\left( \ln \frac{t}{k\, \theta}\right)^{\alpha-1} \Lambda [X_k] ,
\end{equation}
where $m-1<\alpha<m$, and $0<a<\theta$.
\end{Corollary}

For the operator $\, ^C{\cal D}^{t,\alpha}_{a,\eta}$, the analog of Theorem \ref{Theorem-Xt-1} is formulated in the form.

\begin{Theorem} \label{Theorem-Xt-1C}
Let $X(t) \in AC^{m}_{{\cal S},\eta}(a,b)$ 
with $m-1<\alpha \le m$, $a \in (0,\theta)$, 
and $\Lambda [X(t-\epsilon)] \in C^{\infty}(a,b)$.
The integro-differential equation 
\begin{equation} \label{FDE-HT1C}
^C{\cal D}^{t,\alpha}_{a,\eta}[\tau] X(\tau) +
A \, \Lambda [X(t-\epsilon)] \sum^{\infty}_{k=1} \delta \Bigl(\frac{t}{\theta}-k \Bigr)=0 
\end{equation}
has the solution for $t \in (n\theta,(n+1)\theta)$ in the form
\begin{equation} \label{FDE-HT4bC}
X(t) = \sum^{m-1}_{j=0} \frac{C_j^{*}}{\Gamma(j+1)} 
\left( \frac{a}{t}\right)^{\eta}
\left( \ln \frac{t}{a}\right)^{j}
- \frac{A}{\Gamma(\alpha)} 
\sum^{n}_{k=1} \frac{1}{k} 
\left( \frac{k\, \theta}{t}\right)^{\eta} 
\left( \ln \frac{t}{k\, \theta}\right)^{\alpha-1} \Lambda [X(t_k-\epsilon)],
\end{equation}
where $t_k=k\theta$ and
\begin{equation} \label{FTFC-HT-3b-C}
C_j^{*} = \lim_{t \to a+} D^{t,j}_{a,\eta} X(t) , \quad (j=0,...,m-1) .
\end{equation}
\end{Theorem}

{\bf Proof.}
The proof of this theorem is similar to the proof of Theorem \ref{Theorem-Xt-1}. 
The difference between the solutions lies only in the term that contains the initial conditions described by the coefficients $C_j$, $j=1,...,m$. 

\ \ \ $\Box$ \\


\section{Non-local dynamics with parameter $\alpha \in (0,1)$}

Let us consider equation \eqref{FDE-HT1} for $\alpha \in (0,1)$. 
Using the solution of equation \eqref{FDE-HT1}, which is described by Theorem \ref{Theorem-Xt-1}, we can derive the mappings with non-locality in time for $\alpha \in (0,1)$.

\begin{Theorem} \label{Theorem-4-1}
Let $X(t) \in AC^{1}_{{\cal S},\eta}(a,b)$ 
with $0<\alpha \le 1$, $a \in (0,\theta)$, 
and $\Lambda [X(t-\epsilon)] \in C^{\infty}(a,b)$.
The solution of the equation
\begin{equation} \label{MAP-EQ-HT1-1}
{\cal D}^{t,\alpha}_{a,\eta}[\tau] X(\tau) = - A \, \Lambda [X(t-\epsilon)] 
\sum^{\infty}_{k=1} \delta \Bigl(\frac{t}{\theta}-k \Bigr) ,
\end{equation}
for $t=t_{n+1}-\epsilon$, where $t_k=k\theta$, $0<a<\theta$, 
for the variables 
\begin{equation} \label{Def-X-k}
X_{k}=\lim_{\epsilon \rightarrow 0+} X(k\theta-\epsilon),
\quad (k=1,...,n+1)
\end{equation} 
is represented by the non-local mapping
\[
X_{n+1} -X_n = 
\frac{C_1}{a^{\eta -1} \, \theta \, \Gamma(\alpha)} 
{\cal W}_{\alpha,\eta}(n,a/\theta) -
\frac{A}{\Gamma(\alpha)} {\cal V}_{\alpha,\eta}(n+1,n) \, \Lambda [X_n] -
\]
\begin{equation} \label{FDE-HT5map2}
\frac{A}{\Gamma(\alpha)} \sum^{n-1}_{k=1} 
{\cal W}_{\alpha,\eta}(n,k) \, \Lambda [X_k] ,
\end{equation}
where $0<a/\theta<1$, and
\begin{equation} \label{Vam}
{\cal V}_{\alpha,\eta}(x,y) = 
\frac{1}{y} \left( \frac{y}{x} \right)^{\eta} 
\left( \ln \frac{x}{y}\right)^{\alpha-1} ,
\end{equation}
\begin{equation} \label{Wam}
{\cal W}_{\alpha,\eta}(x,y) = 
{\cal V}_{\alpha,\eta}(x+1,y) -{\cal V}_{\alpha,\eta}(x,y) ,
\end{equation}
where $x,y>0$.
\end{Theorem}

{\bf Proof.}
Using Theorem \ref{Theorem-Xt-1}, the solution of equation \eqref{MAP-EQ-HT1-1} is given by expression \eqref{FDE-HT4b} with $\alpha \in (0,1)$ and $m=1$ in the form
\begin{equation} \label{FDE-HT4b1}
X(t) = \frac{C_1\, t^{-\eta}}{\Gamma(\alpha)} 
\left( \ln \frac{t}{a}\right)^{\alpha-1}
- \frac{A}{\Gamma(\alpha)} 
\sum^{n}_{k=1} \frac{1}{k} 
\left( \frac{k\, \theta}{t}\right)^{\eta} 
\left( \ln \frac{t}{k\, \theta}\right)^{\alpha-1} 
\Lambda [X(t_k-\epsilon)] ,
\end{equation}
where $0<a<\theta$, $t_k=k\theta$ and
\begin{equation} \label{FDE-HT6}
C_1 = \lim_{t\to a+} 
t^{\eta}{\cal J}^{t,1-\alpha}_{a,\eta} [\tau] X(\tau) .
\end{equation}

Using $t=(n+1)\theta-\epsilon$, $\epsilon >0$, and the variables 
\begin{equation} 
X_{n+1}=\lim_{\epsilon \rightarrow 0+} X(\theta(n+1)-\epsilon),
\end{equation}
solution \eqref{FDE-HT4b1} gives 
\begin{equation} \label{FDE-HT5}
X_{n+1} = \frac{C_1 \, (n+1)^{-\eta} \, \theta^{-\eta}}{\Gamma(\alpha)} 
\left( \ln \frac{n+1}{a/\theta}\right)^{\alpha-1} -
\frac{A}{\Gamma(\alpha)} \sum^{n}_{k=1} 
\frac{1}{k} \left( \frac{k}{n+1}\right)^{\eta} 
\left( \ln \frac{n+1}{k}\right)^{\alpha-1} \Lambda [X_k] ,
\end{equation}
where $0<a/\theta<1$.
Using $t=n\theta-\epsilon$, where $\epsilon>0$ and $\epsilon \to 0$, we obtain
\begin{equation} \label{FDE-HT5n}
X_{n} = \frac{C_1 \, n^{-\eta} \, \theta^{-\eta}}{\Gamma(\alpha)} 
\left( \ln \frac{n}{a/\theta}\right)^{\alpha-1} -
\frac{A}{\Gamma(\alpha)} \sum^{n-1}_{k=1} 
\frac{1}{k} \left( \frac{k}{n}\right)^{\eta} 
\left( \ln \frac{n}{k}\right)^{\alpha-1} \Lambda [X_k] .
\end{equation}
Subtracting expression \eqref{FDE-HT5n} from equation \eqref{FDE-HT5}, we obtain the mapping
\[
X_{n+1} -X_n = 
\frac{C_1}{\theta^{\eta}\Gamma(\alpha)} \left(
(n+1)^{-\eta} \left( \ln \frac{n+1}{a/\theta}\right)^{\alpha-1} -
n^{-\eta} \left( \ln \frac{n}{a/\theta}\right)^{\alpha-1}
\right) -
\]
\[
\frac{A}{\Gamma(\alpha)} 
\frac{1}{n} \left( \frac{n}{n+1}\right)^{\eta} 
\left( \ln \frac{n+1}{n}\right)^{\alpha-1} \Lambda [X_n] -
\]
\begin{equation} \label{FDE-HT5map}
\frac{A}{\Gamma(\alpha)} \sum^{n-1}_{k=1} 
\left(
\frac{1}{k} \left( \frac{k}{n+1}\right)^{\eta} 
\left( \ln \frac{n+1}{k}\right)^{\alpha-1} -
\frac{1}{k} \left( \frac{k}{n}\right)^{\eta} 
\left( \ln \frac{n}{k}\right)^{\alpha-1} 
\right) \, \Lambda [X_k] .
\end{equation}

Using function \eqref{Vam}, mappings with non-local scaling \eqref{FDE-HT5map} is written as
\[
X_{n+1} -X_n = 
\frac{C_1}{a^{\eta -1} \, \theta \, \Gamma(\alpha)} 
\Bigl( {\cal V}_{\alpha,\eta}(n+1,a/\theta) - {\cal V}_{\alpha,\eta}(n,a/\theta) \Bigr) -
\frac{A}{\Gamma(\alpha)} {\cal V}_{\alpha,\eta}(n+1,n) \, \Lambda [X_n] -
\]
\begin{equation} \label{Eq-Proof-1}
\frac{A}{\Gamma(\alpha)} \sum^{n-1}_{k=1} 
\Bigl( {\cal V}_{\alpha,\eta}(n+1,k) - {\cal V}_{\alpha,\eta}(n,k) \Bigr)
\, \Lambda [X_k] .
\end{equation}
As a result, using \eqref{Wam}, equation \eqref{Eq-Proof-1} gives
\eqref{FDE-HT4b1}.

\ \ \ $\Box$ \\

\begin{Remark}
We should emphasize an important property of mappings with non-locality \eqref{FDE-HT5map2}, namely the fact that mapping \eqref{FDE-HT5map2} does not depend on the kick period $\theta$, if $C_1=0$. 
\end{Remark}

\begin{Remark}
For equation with HFD ($\eta=0$) of the order $\alpha \in (0,1)$, the mappings with non-local scaling leads to \eqref{FDE-HT5map2} with $\eta=0$.
\end{Remark}

\begin{Remark}
For example, if $A \, \Lambda [X]=(1-r) \, X + r \, X^2$, then equation \eqref{FDE-HT5map2} gives a generalization of the logistic mapping, in which non-local scaling in time is taken into account.
\end{Remark}


Using the solution \eqref{FDE-HT4bC} of equation \eqref{FDE-HT1C}, which is described by Theorem \ref{Theorem-Xt-1C}, 
we can derive the mappings with non-local scaling in time from equation \eqref{FDE-HT1C} with Caputo modification of HTFD operator with $\alpha \in (0,1)$.

\begin{Theorem} \label{Theorem-4-2}
Let $X(t) \in AC^{m}_{{\cal S},\eta}(a,b)$ 
with $0<\alpha \le 1$, $a>0$,
and $\Lambda [X(t-\epsilon)] \in C^{\infty}(a,b)$.
The solution of the equation
\begin{equation} \label{MAP-EQ-HT1-1-C}
^C{\cal D}^{t,\alpha}_{a,\eta}[\tau] X(\tau) = - 
A \, \Lambda [X(t-\epsilon)] 
\sum^{\infty}_{k=1} \delta \Bigl(\frac{t}{\theta} - k \Bigr) ,
\end{equation}
for $t=t_{n+1}-\epsilon$, where $t_k=k\theta$, $0<a<\theta$, 
for variables $X_k$ given by $X(k\theta-\epsilon)$ at $\epsilon \rightarrow 0+$ in \eqref{Def-X-k}
is represented by the non-local mapping
\[
X_{n+1} -X_n = a \, C_0^{*} \, {\cal W}_{1,\eta}(n,a/\theta) -
\frac{A}{\Gamma(\alpha)} {\cal V}_{\alpha,\eta}(n+1,n) \, \Lambda [X_n] -
\]
\begin{equation} \label{FDE-HT5map2-C}
\frac{A}{\Gamma(\alpha)} \sum^{n-1}_{k=1} 
{\cal W}_{\alpha,\eta}(n,k) \, \Lambda [X_k] ,
\end{equation}
where $C_0^{*}=X(a+)$, $0<a/\theta<1$, and the function
${\cal V}_{\alpha,\eta}(x,y)$ and ${\cal W}_{\alpha,\eta}(x,y)$ are defined by
\eqref{Vam} and \eqref{Wam} with $x,y>0$.
\end{Theorem}

{\bf Proof.}
The proof of this theorem is similar to the proof of Theorem \ref{Theorem-4-1} by using solution \eqref{FDE-HT4bC} of equation \eqref{FDE-HT1C}, which is described by Theorem \ref{Theorem-Xt-1C}. 

\ \ \ $\Box$ \\

Equation \eqref{FDE-HT5map2-C} defines the one-dimensional non-local mapping, 
which differs from mappings \eqref{FDE-HT5map2} only by the term containing the initial conditions.


\section{Nonlinear dynamics with parameter $\alpha = 1$}

Let us consider equations \eqref{MAP-EQ-HT1-1}, \eqref{MAP-EQ-HT1-1-C} for the integer order $\alpha =1$.
In this case, the mappings are described by the special cases of
Theorems \ref{Theorem-4-1} and \ref{Theorem-4-2}, where the constants $C_1$ and $C_1^{*}$ are
$C_1 = a^{\eta} X(a)$ and $C_1^{*}=X(a)$.

\begin{Corollary}
For $\alpha=1$, equations \eqref{MAP-EQ-HT1-1}, \eqref{MAP-EQ-HT1-1-C} are the first-order differential equation
\begin{equation} 
t^{-\eta+1} \frac{d}{dt} t^{\eta} X(t) + 
A \, \Lambda [X(t-\epsilon)] 
\sum^{\infty}_{k=1} \delta \Bigl(\frac{t}{\theta}-k \Bigr) = 0 
\end{equation}
that leads to mappings \eqref{FDE-HT5map2}, \eqref{FDE-HT5map2-C} with $\alpha=1$ in the form
\[
X_{n+1}= X_n + X(a) \left(\frac{a}{\theta} \right)^{\mu} 
\left( (n+1)^{-\eta} - n^{-\eta} \right) -
\frac{A}{n} \left( \frac{n}{n+1}\right)^{\eta} \Lambda [X_n] -
\]
\begin{equation} \label{Alpha-1-eta}
A \, \left( (n+1)^{-\eta} - n^{-\eta} \right) 
\sum^{n-1}_{k=1} k^{\eta-1} \, \Lambda [X_k] .
\end{equation}
\end{Corollary}

We can see that mapping \eqref{Alpha-1-eta} does not depend on the kick period $\theta$, if $X(a)=0$. 

\begin{Corollary}
For $\alpha=1$ and $\eta=0$, equations \eqref{MAP-EQ-HT1-1}, \eqref{MAP-EQ-HT1-1-C} are the first-order differential equation
\begin{equation} 
t \frac{d}{dt} X(t) + A \, \Lambda [X(t-\epsilon)] 
\sum^{\infty}_{k=1} \delta \Bigl(\frac{t}{\theta}-k \Bigr) = 0 
\end{equation}
that gives the mappings \eqref{FDE-HT5map2}, \eqref{FDE-HT5map2-C} with $\alpha=1$ and $\eta=0$, and ${\cal V}_{1,0}(x,y)=1/y$
have the form
\begin{equation} \label{FDE-HT7-0c}
X_{n+1} = X_{n} - \frac{A}{n} \, \Lambda [X_n] .
\end{equation}
\end{Corollary}

Equation \eqref{FDE-HT7-0c} describes local mapping.
We can see that mapping \eqref{FDE-HT7-0c} does not depend on the kick period, i.e. equation \eqref{FDE-HT7-0c} does not change when the period $\theta$ changes.


\section{Non-local dynamics with parameter $\alpha >1$}

Before deriving non-local mappings for $\alpha> 1$, we will prove some properties of the HTFD operators and define new variables $Y_s$ that can be interpreted as generalized momenta for writing integro-differential equations in the generalized Hamiltonian form. 
The need to define this new variable is due to the fact that the mappings for $\alpha>1$ are $m$-dimensional mappings, where 
$m=[\alpha]+1$ for non-integer values of $\alpha$, and 
$m=\alpha$ for $\alpha \in \mathbb{N}$, i.e. $m-1<\alpha \le m$.

\subsection{New variables $Y_s(t)$}

Let us consider the some properties of the HTFD operators that are described by equations \eqref{DEF-DER-HT} and \eqref{EQ-HT-m}.

\begin{Theorem} \label{Theorem-5-1}
Let $X(t) \in AC^m_{{\cal S},\eta}[a,b]$ 
with $m-1<\alpha \le m$ and $\alpha > 1$, $a>0$. 
Then the following equalities
\begin{equation} \label{DEF-DER-HT-P3}
{\cal D}^{t,1}_{a,\eta} {\cal D}^{t,m-1}_{a,\eta} X(t) = 
{\cal D}^{t,m}_{a,\eta} X (t) ,
\end{equation}
\begin{equation} \label{DEF-DER-HT-P5}
{\cal D}^{t,\alpha}_{a,\eta} [\tau] X (\tau) = 
{\cal D}^{t,1}_{a,\eta} \, 
{\cal D}^{t,\alpha-1}_{a,\eta} [\tau] X (\tau) 
\end{equation}
are satisfied, if $t>a>0$ and $m \in \mathbb{N}$, $m-1>0$.
\end{Theorem}

\textbf{Proof.}
Let us consider the HTFD operator of the order $m \in \mathbb{N}$ that is defined by equation \eqref{EQ-HT-m} in the form
\begin{equation} \label{DEF-DER-HT-P2}
{\cal D}^{t,m}_{a,\eta} X(t) = 
t^{-\eta} \left(t \frac{d}{dt}\right)^m \left( t^{\eta} \, X(t) \right) ,
\end{equation}
where $t>a>0$.
For all $m \in \mathbb{N}$ such that $m-1>0$, we have
\[
{\cal D}^{t,1}_{a,\eta} {\cal D}^{t,m-1}_{a,\eta} X (t) = 
\left( t^{-\eta} \left(t \frac{d}{dt} \right)^1 t^{\eta} \right) \,
\left(t^{-\eta} \left(t \frac{d}{dt}\right)^{m-1} t^{\eta} \right)\, X(t) =
\]
\begin{equation} \label{EQ-62-B}
\left( t^{-\eta} \left(t \frac{d}{dt} \right)^1 
\left(t \frac{d}{dt}\right)^{m-1} t^{\eta} \right)\, X(t) =
\left( t^{-\eta} \left( t \frac{d}{dt}\right)^{m} t^{\eta} \right) \, X(t) 
= {\cal D}^{t,m}_{a,\eta} [\tau] X(\tau) .
\end{equation}
Therefore, the equality
\begin{equation} \label{DEF-DER-HT-P3b}
{\cal D}^{t,1}_{a,\eta} {\cal D}^{t,m-1}_{a,\eta} X(t) = 
{\cal D}^{t,m}_{a,\eta} X(t) 
\end{equation}
holds for $m \in \mathbb{N}$, $m-1>0$.
Using \eqref{DEF-DER-HT-P2}, 
the HTFD operator is written as 
\begin{equation} \label{DEF-DER-HT-P1}
{\cal D}^{t,\alpha}_{a,\eta}[\tau] X(\tau) = 
t^{-\eta} \left(t \frac{d}{dt}\right)^m t^{\eta} 
{\cal J}^{t,m-\alpha}_{a,\eta} [\tau] X(\tau) = 
{\cal D}^{t,m}_{a,\eta} 
{\cal J}^{t,m-\alpha}_{a,\eta} [\tau] X(\tau) .
\end{equation}
Using equations \eqref{EQ-62-B} and \eqref{DEF-DER-HT-P1}
with $\alpha>1$, where $m-1<\alpha \le m$ we have
\[
{\cal D}^{t,\alpha}_{a,\eta}[\tau] X(\tau) = 
{\cal D}^{t,m}_{a,\eta} \,
{\cal J}^{t,m-\alpha}_{a,\eta} [\tau] X(\tau) =
{\cal D}^{t,1}_{a,\eta} \, 
{\cal D}^{t,m-1}_{a,\eta} \,
{\cal J}^{t,m-\alpha}_{a,\eta} [\tau] X(\tau) =
\]
\[
{\cal D}^{t,1}_{a,\eta} \, {\cal D}^{t,m-1}_{a,\eta} \,
{\cal J}^{t,(m-1)-(\alpha-1)}_{a,\eta} [\tau] X(\tau) =
{\cal D}^{t,1}_{a,\eta} \, {\cal D}^{t.\alpha-1}_{a,\eta} [\tau] X(\tau) .
\]
Therefore, the property
\begin{equation} \label{DEF-DER-HT-P5b}
{\cal D}^{t,\alpha}_{a,\eta}[\tau] X(\tau) = 
({\cal D}^{t,1}_{a,\eta} \, {\cal D}^{t,\alpha-1}_{a,\eta} [\tau] X(\tau) ) 
\end{equation}
holds for $\alpha>1$.

\ \ \ $\Box$ \\

\begin{Corollary}
Let $X(t) \in AC^m_{{\cal S},\eta}[a,b]$ with $m-1<\alpha \le m \in \mathbb{N}$. Then the equality
\begin{equation} \label{DEF-DER-HT-P5c}
{\cal D}^{t,\alpha}_{a,\eta}[\tau] X(\tau) = 
{\cal D}^{t,k}_{a,\eta} \, 
{\cal D}^{t,\alpha-k}_{a,\eta} [\tau] X(\tau) 
\end{equation}
holds for $k=1, ..., m-1$ and $t>a>0$.
\end{Corollary}

\textbf{Proof.}
The proof is based on repeated applications of equality 
\eqref{DEF-DER-HT-P5} of Theorem \ref{Theorem-5-1}.

\ \ \ $\Box$ \\

To derive discrete-time non-local mappings from integro-differential equations \eqref{FDE-HT1} of the order $\alpha >1$, we should represent these equations in the Hamiltonian form.
Property \eqref{DEF-DER-HT-P5c} allows us to define new variables by the recurrence relations 
\begin{equation} \label{DEF-DER-HT-P6a}
Y_1(t) = {\cal D}^{t,\alpha-(m-1)}_{a,\eta}[\tau] X(\tau) , 
\end{equation}
\begin{equation} \label{DEF-DER-HT-P6b}
Y_{s+1} (t) = {\cal D}^{t,1}_{a,\eta} \, Y_s(t) ,
\end{equation}
where $s=1,...,m-1$.

Let us define the variables $Y_{s} (t)$.

\begin{Definition}
Let $X(t) \in AC^{s}_{{\cal S}, \eta}$, where $s =1, ..., m-1$,
and $m-1 < \alpha <m$. 
Then the variables $Y_{s}(t)$ are defined as
\begin{equation} \label{DEF-DER-HT-P7}
Y_{s} (t) = {\cal D}^{t,\alpha-m+s}_{a,\eta} [\tau] X(\tau) ,
\end{equation}
where $s=1,...,m-1$, $m-1<\alpha \le m$.
\end{Definition}

Using variables \eqref{DEF-DER-HT-P7}, we formulate the following theorem.


\begin{Theorem} \label{Theorem-5-2}
If $X(t) \in AC^{s}_{{\cal S}, \eta}(a.b)$, 
where $s =1, ..., m-1$, where $m-1<\alpha < m$. 
Then the variables $Y_{s}(t)$ ($s =1, ..., m-1$) take the form
\begin{equation} \label{TH-Ps-NEW}
Y_{s} (t) = 
{\cal D}^{t,s}_{a,\eta} 
{\cal J}^{t,m-\alpha}_{a,\eta} [\tau] X(\tau) =
t^{-\eta} \left(t \frac{d}{dt}\right)^{s} t^{\eta} 
{\cal J}^{t,m-\alpha}_{a,\eta} [\tau] X(\tau) ,
\end{equation}
where $s=1,...,m-1$.
\end{Theorem}

{\bf Proof.}
Using equation \eqref{DEF-DER-HT-P1} and \eqref{DEF-DER-HT-P7}, variables \eqref{DEF-DER-HT-P7} are represented in the form
\begin{equation} \label{DEF-DER-HT-P7s1}
Y_{s} (t) = {\cal D}^{t,\alpha-m+s}_{a,\eta} [\tau] X(\tau) =
{\cal D}^{t,\alpha^{\prime}}_{a,\eta} [\tau] X(\tau) =
t^{-\eta} \left(t \frac{d}{dt}\right)^{m^{\prime}} t^{\eta} 
{\cal J}^{t,m^{\prime}-\alpha^{\prime}}_{a,\eta} [\tau] X(\tau) ,
\end{equation}
where $\alpha^{\prime}=\alpha-m+s$ and 
\[
m^{\prime} = [\alpha^{\prime}]+1=([\alpha]+1)-m+s+1=s, 
\]
\[
m^{\prime}-\alpha^{\prime} = s -(\alpha-m+s)= m-\alpha .
\]
Then we obtain
\begin{equation} \label{DEF-DER-HT-P7s2}
Y_{s} (t) = t^{-\eta} \left(t \frac{d}{dt}\right)^{s} t^{\eta} 
{\cal J}^{t,m-\alpha}_{a,\eta} [\tau] X(\tau) .
\end{equation}

$\ \ \ \Box$ \\


\begin{Theorem} \label{Theorem-5-3}
If $X(t) \in AC^{m-k}_{{\cal S}, \eta}(a,b)$, 
where $k =1, ..., m-1$, $a>0$,
and $m-1 < \alpha <m$, then the initial conditions
\begin{equation} \label{FTFC-HT-3c-0}
C_k = \lim_{t \to a+} \left(t \frac{d}{dt}\right)^{m-k} t^{\eta}
{\cal J}^{t,m-\alpha}_{a,\eta} [\tau] X(\tau) , 
\quad (k=1,...,m) .
\end{equation}
are written through the variables $Y_{s}(t)$ as
\begin{equation} \label{FTFC-HT-3d-0}
C_k =
\lim_{t \to a+} \left( t^{\eta} \, Y_{m-k} (t) \right) , \quad
(k=1,...,m) ,
\end{equation}
where $Y_0(t)=X(t)$.
\end{Theorem}

{\bf Proof.}
The conditions
\begin{equation} \label{FTFC-HT-3c}
C_k= \lim_{t \to a+} \left(t \frac{d}{dt}\right)^{m-k} t^{\eta}
{\cal J}^{t,m-\alpha}_{a,\eta} [\tau] X(\tau) , \quad (k=1,...,m) 
\end{equation}
can be written through the variables $Y_{s}(t)$ by using the following
\[
\left(t \frac{d}{dt}\right)^{m-k} t^{\eta}
{\cal J}^{t,m-\alpha}_{a,\eta}[\tau] X(\tau) =
t^{\eta}
\left( t^{-\eta} \left(t \frac{d}{dt}\right)^{m-k} t^{\eta} \right) {\cal J}^{t,m-\alpha}_{a,\eta} [\tau] X(\tau) =
\]
\[
t^{\eta}
\left( t^{-\eta} \left(t \frac{d}{dt}\right)^{m-k} t^{\eta} \right) {\cal J}^{t,(m-k)-(\alpha-k)}_{a,\eta} [\tau] X(\tau) =
t^{\eta} \, {\cal D}^{t,\alpha-k}_{a,\eta} [\tau] X(\tau) ,
\]
where $k=1,...,m$, and
using Theorem \ref{Theorem-5-2} and equation \eqref{TH-Ps-NEW}, we obtain
\[
\left(t \frac{d}{dt}\right)^{m-k} t^{\eta}
{\cal J}^{t,m-\alpha}_{a,\eta} [\tau] X(\tau) =
t^{\eta}
\left( t^{-\eta} \left(t \frac{d}{dt}\right)^{m-k} t^{\eta} \right) {\cal J}^{t,m-\alpha}_{a,\eta} [\tau] X(\tau) =
t^{\eta} \, Y_{m-k} (t) .
\]
Therefore, we have
\begin{equation} \label{FTFC-HT-3d}
C_k= \lim_{t \to a+} \left( t^{\eta} \, Y_{m-k} (t) \right) , \quad (k=1,...,m) .
\end{equation}

$\ \ \ \Box$ \\


\subsection{Equations and mappings for variables $Y_{s}(t)$}

To derive mappings with non-local scaling from equations 
\begin{equation} \label{MAP-Caputo-HTFD-1}
^C{\cal D}^{t,\alpha}_{a,\eta}[\tau] X(\tau) = - A \, \Lambda [X(t-\epsilon)] 
\sum^{\infty}_{k=1} \delta \Bigl(\frac{t}{\theta}-k \Bigr) ,
\end{equation}
of the order $\alpha>1$, we define the variables $Y_s(t)$ that can be interpreted as the generalized momenta. 
Using the variables $Y_{s}(t)$, we write equation 
\eqref{MAP-Caputo-HTFD-1} with $\alpha>1$ as the following systems of equations

1) For $\alpha \in (m-1,m)$ with $m>2$, equation \eqref{FDE-HT1} with $\alpha \in (m-1,m)$, is written in the generalized Hamiltonian form 
\begin{equation} \label{MAP-EQ-HT-1a}
\begin{cases}
{\cal D}^{t,\alpha-(m-1)}_{a,\eta} [\tau] X(\tau) & = Y_1 (t) , \\
{\cal D}^{t,1}_{a,\eta} Y_s(t) & = Y_{s+1}(t) , 
\quad (s = 1,...,m-2) , \\
{\cal D}^{t,1}_{a,\eta} Y_{m-1} (t) & = - A \, \Lambda [X(t-\epsilon)] 
\displaystyle
\sum^{\infty}_{k=1} \delta \Bigl(\frac{t}{\theta}-k \Bigr) ,
\end{cases}
\end{equation}
where $m-1<\alpha \le m$, $0<a<\theta$. 

2) For $\alpha \in (1,2)$, i.e. $m=2$, equation \eqref{FDE-HT1} is be written a the system 
\begin{equation} \label{MAP-EQ-HT-1b}
\begin{cases}
{\cal D}^{t,\alpha-1}_{a,\eta} [\tau] X(\tau) & = Y_1 (t) , \\
{\cal D}^{t,1}_{a,\eta} Y_1(t) & = - A \, \Lambda [X(t-\epsilon)] 
\displaystyle
\sum^{\infty}_{k=1} \delta \Bigl(\frac{t}{\theta}-k \Bigr) .
\end{cases}
\end{equation}

We will consider equations \eqref{MAP-EQ-HT-1a} and \eqref{MAP-EQ-HT-1b} with the initial conditions
\begin{equation} \label{Init-Cond-Ps1}
\lim_{t \to a+} t^{\eta} \, Y_{m-k}(t) = C_k ,
\quad (k=1,...,m-1) ,
\end{equation}
and
\begin{equation} \label{Init-Cond-Ps2}
\lim_{t \to a+} t^{\eta} 
{\cal J}^{t,m-\alpha}_{a,\eta} [\tau] X(\tau) = C_m .
\end{equation}

Let us prove the following theorem for systems \eqref{MAP-EQ-HT-1a} and \eqref{MAP-EQ-HT-1b}.

\begin{Theorem} \label{Theorem-Ps}
Let $Y_s(t) \in AC^{m-s}_{{\cal S},\eta}(a,b)$, 
$X(t) \in AC^{m}_{{\cal S},\eta}(a,b)$ 
with $m-1<\alpha \le m$, $a>0$,
and $\Lambda [X(t-\epsilon)] \in C^{\infty}(a,b)$.
Equations \eqref{MAP-EQ-HT-1a} and \eqref{MAP-EQ-HT-1b} for $n\theta <t<(n+1)\theta$ with initial conditions \eqref{Init-Cond-Ps1} and \eqref{Init-Cond-Ps2} have the solutions
\[
Y_s(t) = \sum^{m-s}_{j=1} 
\frac{C_j \, t^{-\eta}}{\Gamma(m-s-j+1)} 
\left( \ln \frac{t}{a}\right)^{m-s-j} - 
\]
\begin{equation} \label{FDE-HT4c2}
\frac{A}{\Gamma(m-s)} \sum^{n}_{k=1} \frac{1}{k} 
\left( \frac{k\, \theta}{t}\right)^{\eta} 
\left( \ln \frac{t}{k\, \theta}\right)^{m-s-1} \, \Lambda [X(t_k-\epsilon)],
\quad (s=1,...,m-1) ,
\end{equation}
where $m-1<\alpha \le m$, $0<a<\theta$, $t_k=k\theta$, and
the solutions for $X(t)$ is given as
\begin{equation} \label{FDE-HT4b2}
X(t) = \sum^{m}_{j=1} \frac{C_j \, t^{-\eta}}{\Gamma(\alpha-j+1)} 
\left( \ln \frac{t}{a}\right)^{\alpha-j}
- \frac{A}{\Gamma(\alpha)} 
\sum^{n}_{k=1} \frac{1}{k} 
\left( \frac{k\, \theta}{t}\right)^{\eta} 
\left( \ln \frac{t}{k\, \theta}\right)^{\alpha-1} \Lambda [X(t_k-\epsilon)] ,
\end{equation}
where $0<a<\theta$.
\end{Theorem}

{\bf Proof.}
To get the solutions of integro-differential equations 
\eqref{MAP-EQ-HT-1a} and \eqref{MAP-EQ-HT-1b} for the variables $Y_s(t)$, we will use the equations
\begin{equation} \label{FDE-HT1b-0}
{\cal D}^{t,m-s}_{a,\eta} Y_s(t) +
A \, \Lambda [X(t-\epsilon)] \sum^{\infty}_{k=1} 
\delta \Bigl(\frac{t}{\theta}-k \Bigr)=0 , \quad (s=1,...,m-1)
\end{equation}
and the initial conditions
\begin{equation} \label{Init-Cond-Ps1p}
\lim_{t \to a+} t^{\eta} \, Y_{m-k}(t) = C_k ,
\quad (k=1,...,m-1) ,
\end{equation}
and
\begin{equation} \label{Init-Cond-Ps2p}
\lim_{t \to a+} \left(t^{\eta} 
{\cal J}^{t,m-\alpha}_{a,\eta} [\tau] X(\tau) \right) = C_m ,
\end{equation}
where $m-1<\alpha \le m$.
Using transformations similar to those used in the proof of Theorem \ref{Theorem-Xt-1}, the solutions of 
equation \eqref{FDE-HT1b-0} for $n\theta <t<(n+1)\theta$ with initial conditions \eqref{Init-Cond-Ps1p} and \eqref{Init-Cond-Ps2p} can be written in form \eqref{FDE-HT4c2}.
The solution for $X(t)$ is described by the solutions of equation \eqref{FDE-HT1} in Theorem \ref{Theorem-Xt-1}.

$\ \ \ \Box$ \\


We will use solutions \eqref{FDE-HT4c2} and \eqref{FDE-HT4b2} to get the mappings with non-local scaling in time for $\alpha>1$.

\begin{Theorem} \label{Theorem-Pn}
Let $X(t) \in AC^{m}_{{\cal S},\eta}(a,b)$, 
$Y_s(t) \in AC^{m-s}_{{\cal S},\eta}(a,b)$ 
with $m-1<\alpha \le m$, $a>0$,
and $\Lambda [X(t-\epsilon)] \in C^{\infty}(a,b)$.
The solution of equations \eqref{MAP-EQ-HT-1a} and \eqref{MAP-EQ-HT-1b} for $n\theta <t<(n+1)\theta$ with initial conditions \eqref{Init-Cond-Ps1} and \eqref{Init-Cond-Ps2} for the variables 
\begin{equation} 
Y_{s,k}=\lim_{\epsilon \rightarrow 0+} Y_s(k\theta-\epsilon),
\end{equation}
\begin{equation} 
X_{k}=\lim_{\epsilon \rightarrow 0+} X(k\theta-\epsilon) ,
\end{equation}
where $k=1,...,n+1$, are written by the mappings 
\[
Y_{s,n+1} - Y_{s,n} = \sum^{m-s}_{j=1} 
\frac{C_j}{a^{\eta-1} \, \theta \, \Gamma(m-s-j+1)} 
{\cal W}_{m-s-j+1,\eta}(n,a/\theta) -
\]
\begin{equation} \label{FDE-NEW-3}
\frac{A}{\Gamma(m-s)} \frac{1}{n} 
{\cal V}_{m-s,\eta}(n+1,n)\, \Lambda [X_n] -
\frac{A}{\Gamma(m-s)} \sum^{n-1}_{k=1} 
{\cal W}_{m-s,\eta}(n,k) \, \Lambda [X_k] ,
\end{equation}
where $s=1,...,m-1$, and
\[
X_{n+1} -X_n = 
\sum^{m}_{j=1}
\frac{C_j}{a^{\eta -1} \, \theta \, \Gamma(\alpha-j+1)} 
{\cal W}_{\alpha-j+1,\eta}(n,a/\theta) -
\]
\begin{equation} \label{FDE-NEW-3X}
\frac{A}{\Gamma(\alpha)} {\cal V}_{\alpha,\eta}(n+1,n) \, \Lambda [X_n] -
\frac{A}{\Gamma(\alpha)} \sum^{n-1}_{k=1} 
{\cal W}_{\alpha,\eta}(n,k) \, \Lambda [X_k] ,
\end{equation}
where $0<a/\theta<1$, and 
\begin{equation} \label{Vam2}
{\cal V}_{\alpha,\eta}(x,y) = 
\frac{1}{y} \left( \frac{y}{x} \right)^{\eta} 
\left( \ln \frac{x}{y}\right)^{\alpha-1} , 
\end{equation}
\begin{equation} \label{Wam2}
{\cal W}_{\alpha,\eta}(x,y) = 
{\cal V}_{\alpha,\eta}(x+1,y) -{\cal V}_{\alpha,\eta}(x,y) ,
\end{equation}
where $x,y>0$.
\end{Theorem}

{\bf Proof.}
For $m-1<\alpha<m$, using $t=(n+1)\theta-\epsilon$, $\epsilon >0$, and 
\begin{equation} 
Y_{s,n+1}=\lim_{\epsilon \rightarrow 0+} Y_s((n+1)\theta-\epsilon),
\end{equation}
solution \eqref{FDE-HT4c2} is written by the equation 
\[
Y_{s,n+1} = \sum^{m-s}_{j=1} 
\frac{C_j \, ((n+1)\theta)^{-\eta}}{\Gamma(m-s-j+1)} 
\left( \ln \frac{(n+1)\theta}{a}\right)^{m-s-j} - 
\]
\begin{equation} \label{FDE-NEW-1}
\frac{A}{\Gamma(m-s)} \sum^{n}_{k=1} \frac{1}{k} 
\left( \frac{k}{n+1}\right)^{\eta} 
\left( \ln \frac{n+1}{k}\right)^{m-s-1} \, \Lambda [X_k] .
\end{equation}
Using expression \eqref{FDE-HT4c2} for $t=n\theta-\epsilon$, where $\epsilon>0$ and $\epsilon \to 0$, in the form
\[
Y_{s,n} = \sum^{m-s}_{j=1} 
\frac{C_j \, (n\theta)^{-\eta}}{\Gamma(m-s-j+1)} 
\left( \ln \frac{n\theta}{a}\right)^{m-s-j} - 
\]
\begin{equation} \label{FDE-NEW-2}
\frac{A}{\Gamma(m-s)} \sum^{n-1}_{k=1} \frac{1}{k} 
\left( \frac{k}{n}\right)^{\eta} 
\left( \ln \frac{n}{k}\right)^{m-s-1} \, \Lambda [X_k] .
\end{equation}
Subtracting \eqref{FDE-NEW-2} from \eqref{FDE-NEW-1}, we obtain
\[
Y_{s,n+1} -Y_{s,n} = \sum^{m-s}_{j=1} 
\frac{C_j}{\theta^{\eta}\Gamma(m-s-j+1)} 
\Bigl( (n+1)^{-\eta}
\left( \ln \frac{n+1}{a/\theta}\right)^{m-s-j} - 
n^{-\eta}
\left( \ln \frac{n}{a/\theta}\right)^{m-s-j} 
\Bigr) -
\]
\[
\frac{A}{\Gamma(m-s)} \frac{1}{n} 
\left( \frac{n}{n+1}\right)^{\eta} 
\left( \ln \frac{n+1}{n}\right)^{m-s-1} \, \Lambda [X_n] -
\]
\begin{equation} \label{Equation-109}
\frac{A}{\Gamma(m-s)} \sum^{n-1}_{k=1} 
\frac{1}{k} \Bigl(
\left( \frac{k}{n+1}\right)^{\eta} 
\left( \ln \frac{n+1}{k}\right)^{m-s-1} -
\left( \frac{k}{n}\right)^{\eta} 
\left( \ln \frac{n}{k}\right)^{m-s-1} 
\Bigr) \, \Lambda [X_k] .
\end{equation}
Using the function
\begin{equation} \label{Vam3}
{\cal V}_{\alpha,\eta}(x,y) = 
\frac{1}{y} \left( \frac{y}{x} \right)^{\eta} 
\left( \ln \frac{x}{y}\right)^{\alpha-1} ,
\end{equation}
the mappings \eqref{Equation-109} are presented as
\[
Y_{s,n+1} - Y_{s,n} = \sum^{m-s}_{j=1} 
\frac{C_j}{a^{\eta-1} \, \theta \, \Gamma(m-s-j+1)} 
\Bigl( 
{\cal V}_{m-s-j+1,\eta}(n+1,a/\theta) -
{\cal V}_{m-s-j+1,\eta}(n,a/\theta)
 \Bigr) -
\]
\begin{equation} \label{FDE-NEW-3P}
\frac{A}{\Gamma(m-s)} \frac{1}{n} 
{\cal V}_{m-s,\eta}(n+1,n)\, \Lambda [X_n] -
\frac{A}{\Gamma(m-s)} \sum^{n-1}_{k=1} 
\Bigl(
{\cal V}_{m-s,\eta}(n+1,k) - {\cal V}_{m-s,\eta}(n,k)
\Bigr) \, \Lambda [X_k] ,
\end{equation}
where $s=1,...,m-1$.
Using transformations similar to those used in the proof of Theorem \ref{Theorem-5-1}, we derive
\[
X_{n+1} -X_n = \sum^{m}_{j=1}
\frac{C_j}{a^{\eta -1} \, \theta \, \Gamma(\alpha-j+1)} 
\Bigl( {\cal V}_{\alpha-j+1,\eta}(n+1,a/\theta) - {\cal V}_{\alpha-j+1,\eta}(n,a/\theta) \Bigr) -
\]
\begin{equation} 
\frac{A}{\Gamma(\alpha)} {\cal V}_{\alpha,\eta}(n+1,n) \, 
\Lambda [X_n] -
\frac{A}{\Gamma(\alpha)} \sum^{n-1}_{k=1} 
\Bigl( {\cal V}_{\alpha,\eta}(n+1,k) - {\cal V}_{\alpha,\eta}(n,k) \Bigr)
\, \Lambda [X_k] ,
\end{equation}
where $0<a/\theta<1$, and 
\begin{equation} \label{not1b}
X_{k}=\lim_{\epsilon \rightarrow 0+} X(k\theta-\epsilon) .
\end{equation}
Then, using function \eqref{Wam2}, we obtain equations \eqref{FDE-NEW-3} and \eqref{FDE-NEW-3X}.

$\ \ \ \Box$ \\

Equations \eqref{FDE-NEW-3} and \eqref{FDE-NEW-3X} define the $m$-dimensional mapping with non-local scaling in time.


\subsection{Equations and mappings for Caputo modifications}

To derive mappings with non-local scaling from equations \eqref{Cauchy-HT-1C} of the order $\alpha>1$, we should consider the Hamiltonian form of this equation with the variables $Y_s^*(t)$
that can be interpreted as the generalized momenta. 

\begin{Definition}
Let $X(t) \in AC^{s}_{{\cal S},\eta}$, where $s=0,...,m-1$, 
$m-1<\alpha \le m$. Then the variables $Y_s^*(t)$ are defined as
\begin{equation} \label{Ps-Caputo}
Y_s^*(t) = D^{t,s}_{a,\eta} X(t) ,
\end{equation} 
where $Y_0^*(t)=X(t)$.
\end{Definition}

Using variables \eqref{Ps-Caputo}, the equation of the order $\alpha>1$ with periodic kicks
\begin{equation} \label{Caputo-HT-FDE-1}
^C{\cal D}^{t,\alpha}_{a,\eta} [\tau] X(\tau) = 
- A \, \Lambda [X(t-\epsilon)] 
\sum^{\infty}_{k=1} \delta \Bigl(\frac{t}{\theta}-k \Bigr) ,
\end{equation}
will be written as the following systems of equations with $Y_s^*(t)$.

1) For $\alpha \in (m-1,m)$ with $m>2$, equation \eqref{Caputo-HT-FDE-1} is presented in the form 
\begin{equation} \label{Caputo-System-1}
\begin{cases}
{\cal D}^{t,1}_{a,\eta} Y_s^{*}(t) & = Y_{s+1}^{*}(t) , 
\quad (s = 0,...,m-2) , \\
{\cal D}^{t,\alpha-(m-1)}_{a,\eta}[\tau] Y_{m-1}^{*} (\tau) & 
= - A \, \Lambda [X(t-\epsilon)] 
\displaystyle
\sum^{\infty}_{k=1} \delta \Bigl(\frac{t}{\theta}-k \Bigr) ,
\end{cases}
\end{equation}
where $0<a<\theta$ and $Y_0^{*}(t)=X(t)$. 

2) For $\alpha \in (1,2)$, equation \eqref{Caputo-HT-FDE-1} is written in the form 
\begin{equation} \label{Caputo-System-2}
\begin{cases}
{\cal D}^{t,1}_{a,\eta} X(t) & = Y_1^{*} (t) , \\
{\cal D}^{t,\alpha-1}_{a,\eta} [\tau] Y_1^{*}(\tau) & = 
- A \, \Lambda [X(t-\epsilon)] 
\displaystyle
\sum^{\infty}_{k=1} \delta \Bigl(\frac{t}{\theta}-k \Bigr) .
\end{cases}
\end{equation}

For equation \eqref{Caputo-HT-FDE-1}, we write
the initial conditions through the variables $Y_{k}^{*}(t)$
to consider systems \eqref{Caputo-System-1} and \eqref{Caputo-System-2} with the conditions
\begin{equation} \label{Caputo-Init-Cond-Ps1}
\lim_{t \to a+} Y_{k}^{*}(t) = C_k ,
\quad (k=0,...,m-1) .
\end{equation}

For these systems \eqref{Caputo-System-1} and \eqref{Caputo-System-2}, we formulate the following theorem.

\begin{Theorem} \label{Theorem-Caputo-Ps-1}
Let $X(t) \in AC^{m}_{{\cal S},\eta}(a,b)$, 
$Y_s(t) \in AC^{m-s}_{{\cal S},\eta}(a,b)$ 
with $m-1<\alpha \le m$, $a>0$,
and $\Lambda [X(t-\epsilon)] \in C^{\infty}(a,b)$.
Equations \eqref{Caputo-System-1} and \eqref{Caputo-System-2} for $n\theta <t<(n+1)\theta$ with initial conditions \eqref{Caputo-Init-Cond-Ps1} have the solutions
\[
Y_s^{*}(t) = \sum^{(m-s)-1}_{j=0} 
\frac{C_j^{*}}{\Gamma(j+1)} 
\left( \frac{a}{t}\right)^{\eta}
\left( \ln \frac{t}{a}\right)^{j} - 
\]
\begin{equation} \label{Caputo-HT-FDE-2}
\frac{A}{\Gamma(m-s)} \sum^{n}_{k=1} \frac{1}{k} 
\left( \frac{k\, \theta}{t}\right)^{\eta} 
\left( \ln \frac{t}{k\, \theta}\right)^{m-s-1} \, 
\Lambda [X(t_k-\epsilon)] ,
\quad (s=0,...,m-1) ,
\end{equation}
where $m-1<\alpha \le m$, $0<a<\theta$, $t_k=k\theta$, and
$Y_0^{*}(t)=X(t)$.
\end{Theorem}

{\bf Proof.}
An application the Caputo modification of HTFD operator to $X(t)$ will be written through $Y_s^{*}(t)$ by using the transformations
\[
^C{\cal D}^{t,\alpha}_{a,\eta} [\tau] X(\tau) =
{\cal J}^{t,m-\alpha}_{a,\eta} [\tau] 
{\cal D}^{\tau,m}_{a,\eta} X(\tau) =
\] 
\[
{\cal J}^{t,(m-s)-(\alpha-s)}_{a,\eta} [\tau] 
{\cal D}^{\tau,m-s}_{a,\eta}
{\cal D}^{\tau,s}_{a,\eta} X(\tau) =
{\cal J}^{t,(m-s)-(\alpha-s)}_{a,\eta} [\tau] 
{\cal D}^{\tau,m-s}_{a,\eta} Y_s^{*}(\tau) =
\]
\begin{equation}
^C{\cal D}^{t,\alpha-s}_{a,\eta} [\tau] Y_s^{*}(\tau) ,
\end{equation}
where $m-1 <\alpha <m$, $s=0,..., m-1$, $Y_0^{*}(t)=X(t)$. 
This allows us rewrite equation \eqref{Caputo-HT-FDE-1} with $\alpha>1$ through the variables $Y_s^{*}(t)$. 

To get the solutions of systems \eqref{Caputo-System-1} and \eqref{Caputo-System-2} for the variables $Y_s^{*}(t)$, we will use equations \eqref{Caputo-HT-FDE-1}: in the form
\begin{equation} \label{Caputo-HT-FDE-3}
^C{\cal D}^{t,\alpha-s}_{a,\eta} [\tau] Y_s^{*}(\tau) +
A \, \Lambda [X(t-\epsilon)] \sum^{\infty}_{k=1} 
\delta \Bigl(\frac{t}{\theta}-k \Bigr)=0 , \quad (s=1,...,m-1)
\end{equation}
and the conditions
\begin{equation} \label{Caputo-Init-Cond-Ps2}
\lim_{t \to a+} \, Y_{k}^{*}(t) = C_k^{*} ,
\quad (k=0,...,m-1) .
\end{equation}
Using transformations similar to those used in the proof of Theorem \ref{Theorem-Xt-1C}, the solutions of equation \eqref{Caputo-HT-FDE-3} for $n\theta <t<(n+1)\theta$ with initial conditions \eqref{Caputo-Init-Cond-Ps2} are written by \eqref{Caputo-HT-FDE-2}.

$\ \ \ \Box$ \\


Solutions \eqref{Caputo-HT-FDE-2} will be used to derive the mappings with non-local scaling for $\alpha>1$.

\begin{Theorem} \label{Theorem-Caputo-Ps-2}
Let $X(t) \in AC^{m}_{{\cal S},\eta}(a,b)$, 
$Y_s(t) \in AC^{m-s}_{{\cal S},\eta}(a,b)$ 
with $m-1<\alpha \le m$, $a>0$,
and $\Lambda [X(t-\epsilon)] \in C^{\infty}(a,b)$.
The solution of equations \eqref{Caputo-System-1} and \eqref{Caputo-System-2} for $n\theta <t<(n+1)\theta$ with conditions \eqref{Caputo-Init-Cond-Ps1} for the variables 
\begin{equation} 
Y_{s,k}^{*} = 
\lim_{\epsilon \rightarrow a+} Y_s^{*}(k\theta-\epsilon),
\end{equation}
where $k=1,...,n+1$, and $Y_{0,k}^{*} = X_k$, are written by the mappings 
\[
Y_{s,n+1}^{*} - Y_{s,n}^{*} = \sum^{(m-s)-1}_{j=0} 
\frac{a\, C_j^{*}}{\Gamma(j+1)} {\cal W}_{j+1,\eta}(n,a/\theta) -
\]
\begin{equation} \label{FDE-NEW-Caputo}
\frac{A}{\Gamma(m-s)} \frac{1}{n} 
{\cal V}_{m-s,\eta}(n+1,n)\, \Lambda [X_n] -
\frac{A}{\Gamma(m-s)} \sum^{n-1}_{k=1} 
{\cal W}_{m-s,\eta}(n,k) \, \Lambda [X_k] ,
\end{equation}
where $s=0,...,m-1$, $Y_{0,n}^{*}=X_n$, $0<a/\theta<1$, and 
\begin{equation} \label{Vam2b}
{\cal V}_{\alpha,\eta}(x,y) = 
\frac{1}{y} \left( \frac{y}{x} \right)^{\eta} 
\left( \ln \frac{x}{y}\right)^{\alpha-1} , 
\end{equation}
\begin{equation} \label{Wam2b}
{\cal W}_{\alpha,\eta}(x,y) = 
{\cal V}_{\alpha,\eta}(x+1,y) -{\cal V}_{\alpha,\eta}(x,y) ,
\end{equation}
where $x,y>0$.
\end{Theorem}

{\bf Proof.}
The proof of this theorem is similar to the proof of Theorem \ref{Theorem-Pn}.
Considering solutions \eqref{Caputo-HT-FDE-2} for
$t=(n+1)\theta-\epsilon$, and $t=n\theta-\epsilon$ 
at $\epsilon \to 0$, and then subtracting the second expression from first one, we obtain non-local mapping \eqref{FDE-NEW-Caputo}.

Equation \eqref{FDE-NEW-Caputo} defines the $m$-dimensional mapping with non-local scaling, where $m=[\alpha]+1$ if $\alpha$ is non-integer.


\begin{Remark}
Equations \eqref{FDE-NEW-Caputo} and \eqref{FDE-NEW-3}, \eqref{FDE-NEW-3X} describe the fractional dynamics with non-local scaling (dilation) in the discrete time approach.
We can see that mappings \eqref{FDE-NEW-Caputo} and \eqref{FDE-NEW-3}, \eqref{FDE-NEW-3X} with zero initial conditions 
($C_j^{*}=0$ and $C_j=0$ for $j=0, ... ,m-1$) do not depend on the period of kicks $\theta$. 
These mappings describe exact solutions of integro-differential equations with HTFD operators of order $\alpha$ for discrete time points without approximation. 
Non-local mappings \eqref{FDE-NEW-Caputo} and \eqref{FDE-NEW-3}, \eqref{FDE-NEW-3X} can be used in theory of chaos and nonlinear dynamics for modeling new type of chaotic behavior of various processes that are characterized by time-scale invariance. 
\end{Remark} 

\begin{Remark}
Nonlinear equations \eqref{FDE-NEW-Caputo} and \eqref{FDE-NEW-3}, \eqref{FDE-NEW-3X} can be called the universal mappings with non-local scaling. \\
If $\Lambda [X]=-X$, then linear equations \eqref{FDE-NEW-Caputo} and \eqref{FDE-NEW-3}, \eqref{FDE-NEW-3X} give generalizations of the Anosov-type system, where non-localily are taken into account. \\
If $\Lambda [X]=\, (1-r) \, X + r \, X^2$, then nonlinear equations \eqref{FDE-NEW-Caputo} and \eqref{FDE-NEW-3}, \eqref{FDE-NEW-3X} are the logistic mappings with Hadamard type non-locality in time. \\
For $\Lambda [X]=\sin (X)$, nonlinear equations \eqref{FDE-NEW-Caputo} and \eqref{FDE-NEW-3}, \eqref{FDE-NEW-3X} 
are generalizations of the Chirikov-Taylor mapping \cite{Chirikov} with non-local scaling in time, where non-local scaling is taken into account.
\end{Remark} 


\section{Conclusion}

In this article, to consider fractional dynamics with non-local scaling (dilation) in the discrete and continuous time approach, we use a generalization of the Hadamard fractional operators that are proposed in \cite{BKT2002-1,Kilbas2001}
and called the Hadamard type operators. 
These operators can be interpreted as non-local scaling (dilation) differential operators.
Non-linear integro-differential equations with the Hadamard type operators of non-integer orders are suggested and exact solutions of these equations are obtained for all positive real values of orders.

To derive the discrete-time fractional dynamics with non-local scaling, we used the approach first proposed in 
\cite{Tarasov-Zaslavsky,Tarasov-Map1,Tarasov-Map2} 
and then used in \cite{Springer2010}, \cite{Tarasov-Map3,TT-Logistic,Entropy2021,Mathematics2021,MMAS2021,Mathematics2021-2}. 
This approach allows us to derive discrete-tome mapping with non-locality in time from equations with Hadamard type differential operators of non-integer orders without approximation. 
The mappings with non-local scaling are exact solutions of nonlinear FDE with periodic kicks at fixed time points.
The proposed mappings describe the fractional dynamics with non-local scaling in discrete time approach.
It should be emphasized that these mappings describe the exact solution of integro-differential equations \eqref{FDE-HT1} of an arbitrary positive order without an approximation. 

We assume that non-local mappings \eqref{FDE-NEW-3} and \eqref{FDE-NEW-3X} can be used for modeling a new type of chaotic dynamics that is characterized by time-scale invariance in physics \cite{TS-1,TS-2,TS-3}, economics \cite{TS-4,TS-5} and, biology \cite{TS-6,TS-7,TS-8}. 
The proposed approach can also be applied to describe nonlinear dynamics with non-local scaling in space.

Computer simulations of mapping with non-locality in the form of power-law non-locality are described in \cite{Tarasov-Map3,TT-Edelman1,TT-Edelman2}, \cite{Edelman1,Edelman2,Edelman3,Edelman5}, 
\cite{Edelman7,Edelman-Handbook-2,Edelman-Handbook-4,Edelman2021}, where new types of chaotic behavior and new kinds of attractors have been found. 
However, all these mappings corresponded to the power-law non-locality in time and depended on the kick period $T$.

We expect that the nonlinear systems with the non-local scaling in time exhibit a new type of chaotic motion and new type fractional chaotic attractors that can be characterized by scaling invariance.
The proposed mapping with non-local scaling can be used to study a new type of regular and strange attractors for the fractional dynamics of physical, economic and biological processes by taking into account their scaling properties.




\begin{thebibliography}{100}


\bibitem{Barut} Barut A.O., Raczka R., \textit{Theory of Group Representations and Applications}. Second revised edition. Warszawa: PWN-Polish Scientific Publishers, 1980. 717 pages. ISBN: 83-01-02716-9 

\bibitem{EPJPlus2020} Tarasov, V.E.,
Extension of relativistic mechanics by maximum symmetry group of Maxwell equations. 
\textit{The European Physical Journal Plus}. \textbf{2020}. vol.135. No.4. Article number: 376. 
DOI: 10.1140/epjp/s13360-020-00376-x


\bibitem{FC1} Samko, S.G.; Kilbas, A.A.; Marichev, O.I., 
\textit{Fractional Integrals and Derivatives Theory and Applications}. New York: Gordon and Breach, 1993. 1006 pages. ISBN: 9782881248641


\bibitem{Springer2010} Tarasov, V.E., \textit{Fractional Dynamics: Applications of Fractional Calculus to Dynamics of Particles, Fields and Media}. New York: Springer, 2010. 505 pages. DOI: 10.1007/978-3-642-14003-7


\bibitem{FC2} Kiryakova, V., \textit{Generalized Fractional Calculus and Applications}. New York: Longman and J. Wiley, 1994. 360 pages. ISBN: 9780582219779

\bibitem{FC3} Podlubny, I. \textit{Fractional Differential Equations}. San Diego: Academic Press, 1998. 340 pages.
ISBN: 978-0125588409 

\bibitem{FC4} Kilbas, A.A.; Srivastava, H.M.; Trujillo J.J., 
\textit{Theory and Applications of Fractional Differential Equations}. Amsterdam: Elsevier, 2006. 540 pages.
ISBN: 978-0444518323 


\bibitem{FC5} Diethelm, K., \textit{The Analysis of Fractional Differential Equations: An Application-Oriented Exposition Using Differential Operators of Caputo Type}. Berlin: Springer-Verlag, 2010. 247 pages. DOI: 10.1007/978-3-642-14574-2

\bibitem{Handbook2019-1} Kochubei, A.; Luchko, Yu.,
{Handbook of Fractional Calculus with Applications. Volume 1. Basic Theory.} Berlin, Boston: Walter de Gruyter GmbH, 2019. 481 pages. ISBN: 978-3-11-057081-6 
DOI: 10.1515/9783110571622

\bibitem{Handbook2019-2} Kochubei, A.; Luchko, Yu.,
{Handbook of Fractional Calculus with Applications. Volume 2. Fractional Differential Equations.} Berlin, Boston: Walter de Gruyter GmbH, 2019. 519 pages. ISBN: 978-3-11-057082-3
DOI: 10.1515/9783110571660


\bibitem{Handbook2019-4} Tarasov, V.E., \textit{Handbook of Fractional Calculus with Applications. Volume 4. Application in Physics. Part A.} Berlin, Boston: Walter de Gruyter GmbH, 2019. 306 pages. ISBN: 978-3-11-057088-5 
DOI: 10.1515/9783110571707 

\bibitem{Handbook2019-5} Tarasov, V.E., \textit{Handbook of Fractional Calculus with Applications. Volume 5. Application in Physics. Part B.} Berlin, Boston: Walter de Gruyter GmbH, 2019. 319 pages. ISBN: 978-3-11-057089-2 
DOI: 10.1515/9783110571721

\bibitem{BOOK-MDPI-2020} Tarasov, V.E., \textit{Mathematical Economics: Application of Fractional Calculus}. Basel, Beijing: MDPI, 2020. 278 pages.
ISBN: 978-3-03936-118-2; 978-3-03936-119-9 
DOI: 10.3390/books978-3-03936-119-9
URL: https://www.mdpi.com/books/pdfview/book/2350

\bibitem{BOOK-DG-2021} Tarasov, V.E.; Tarasova, V.V. \textit{Economic Dynamics with Memory: Fractional Calculus Approach}. Berlin, Boston: De Gruyter, 2021. 602 pages.
ISBN: 978-3-11-062460-1 DOI 10.1515/9783110627459


\bibitem{AANT2017} Ahmad, B.; Alsaedi, A.; Ntouyas, S.K.; Tariboon, J., 
\textit{Hadamard-Type Fractional Differential Equations, Inclusions and Inequalities}. 
Cham, Switzerland: Springer International Publishing, 2017. 
ISBN: 978-3-319-52140-4 DOI: 10.1007/978-3-319-52141-1

\bibitem{Hadamard1892}
Hadamard, J., Essay on the study of functions given by their Taylor expansion. (Essai sur letude des fonctions donnees par leur developpement de Taylor). 
\textit{Journal de Mathematiques Pures et Appliquees}. 
\textbf{1892}. Vol.8. No.4. P.101-186.

\bibitem{BKT2002-1}
Butzer, P.L.; Kilbas, A. A.; Trujillo, J.J., 
Fractional calculus in the Mellin setting and Hadamard-type fractional integrals. 
\textit{Journal of Mathematical Analysis and Applications}. 
\textbf{2002}. Vol.269. No.1. P.1-27.
DOI: 10.1016/S0022-247X(02)00001-X

\bibitem{Kilbas2001}
Kilbas, A.A., Hadamard-type fractional calculus. 
\textit{Journal of the Korean Mathematical Society}. 
\textbf{2001}. Vol.38. No.6. P.1191-1204.

\bibitem{KilbasTitjura2002}
Kilbas, A.A.; Titjura, A.A., 
Hadamard-type fractional integrals and derivatives. 
\textit{Trudy Instituta Matematiki NAN Belarusi} (Minsk). 
\textbf{2002}. Vol.11. P.79-87.
URL: http://www.koreascience.or.kr/article/JAKO200111920819409.page

\bibitem{BKT2002-2}
Butzer, P.L.; Kilbas, A.A.; Trujillo, J.J., 
Compositions of Hadamard-type fractional integration operators and the semigroup property. 
\textit{Journal of Mathematical Analysis and Applications}. 
\textbf{2002}. Vol.269. No.2. P.387-400.
DOI: 10.1016/S0022-247X(02)00049-5

\bibitem{BKT2002-3}
Butzer, P.L.; Kilbas, A.A.; Trujillo, J.J., 
Mellin transform analysis and integration by parts for Hadamard-type fractional integrals. 
\textit{Journal of Mathematical Analysis and Applications}. 
\textbf{2002}. Vol.270. No.1. P.1-15. 
DOI: 10.1016/S0022-247X(02)00066-5

\bibitem{KMT2003}
Kilbas, A.A.; Marzan, S.A.; Tityura, A.A., 
Hadamard-type fractional integrals and derivatives and differential equations of fractional order.
\textit{Doklady Akademii Nauk}. \textbf{2003}. Vol.389. No.6. P.734-738. [in Russian] 
URL: https://www.elibrary.ru/item.asp?id=17335599

\bibitem{KMT2003E}
Kilbas, A.A.; Marzan, S.A.; Tityura, A.A., 
Hadamard-type fractional integrals and derivatives and differential equations of fractional order.
\textit{Doklady Mathematics}. \textbf{2003}. Vol.67. No.2. P.263-267.
URL: https://www.elibrary.ru/item.asp?id=27822330


\bibitem{Jarad2012}
Jarad, F.; Abdeljawad, T.; Baleanu, D.,
Caputo-type modification of the Hadamard fractional derivatives. 
\textit{Advances in Difference Equations}. 2012. Vol.2012. Article number: 142. DOI: 10.1186/1687-1847-2012-142 

\bibitem{Jarad2014}
Gambo, Y.Y.; Jarad, F.; Baleanu, D.; Abdeljawad, T., 
On Caputo modification of the Hadamard fractional derivatives. 
\textit{Advances in Difference Equations}. \textbf{2014}. Vol.2014. Article number: 10. DOI: 10.1186/1687-1847-2014-10

\bibitem{Zafar2020}
Zafar, R.; Rehman, M.U.; Shams, M.,
On Caputo modification of Hadamard-type fractional derivative and fractional Taylor series.
\textit{Advances in Difference Equations}. 2020. Vol.2020. Article number: 219. DOI: 10.1186/s13662-020-02658-1

\bibitem{ChangpinLi} 
Ma, L.; Li, Ch., On Hadamard fractional calculus. 
\textit{Fractals}. \textbf{2017}. Vol.25. No.3. Article number: 1750033.
DOI: 10.1142/S0218348X17500335

\bibitem{Garra2013} Garra, R.; Polito, F., On some operators involving Hadamard derivatives. \textit{Integral Transforms and Special Functions}. \textbf{2013}. Vol.24. No.10. P.773-782. DOI: 10.1080/10652469.2012.756875

\bibitem{Garra2018} Garra, R.; Orsingher, E.; Polito, F.
A note on Hadamard fractional differential
equations with varying coefficients and their
applications in probability.
\textit{Mathematics}. \textbf{2018}. Vol.6. No.1. Article number: 4. DOI: 10.3390/math6010004



\bibitem{SUZ} Sagdeev, R.Z.; Usikov, D.A.; Zaslavsky, G.M., 
\textit{Nonlinear Physics. From the Pendulum to Turbulence and Chaos}. New York: Harwood Academic, 1988. 656 pages.
ISBN:	3718648326, 9783718648320

\bibitem{Zaslavsky2} Zaslavsky, G.M., 
\textit{Hamiltonian Chaos and Fractional Dynamics}.
Oxford, New York: Oxford University Press, 2005. 421 pages.
ISBN: 0199535485 

\bibitem{Chirikov} Chirikov, B.V., 
A universal instability of many dimensional oscillator systems
\textit{Physics Reports}. \textbf{1979}. Vol.52. No.5. P.263-379.
DOI: 10.1016/0370-1573(79)90023-1

\bibitem{Schuster} Schuster, H.G., 
\textit{Deterministic Chaos. An Introduction}.
3rd edition. Weinheim: Wiley-VCH, 1995. 
320 pages. ISBN: 978-3527293155 


\bibitem{Tarasov-Zaslavsky} Tarasov, V.E.; Zaslavsky G.M.,
Fractional equations of kicked systems and discrete maps.
\textit{Journal of Physics A}. \textbf{2008}. Vol.41. No.43. Article number: 435101. 16 pages.
DOI: 10.1088/1751-8113/41/43/435101 (arXiv:1107.3953) 

\bibitem{Tarasov-Map1} Tarasov, V.E., Differential equations with fractional derivative and universal map with memory. \textit{Journal of Physics A}. \textbf{2009}. Vol.42. No.46. Article number: 465102. DOI: 10.1088/1751-8113/42/46/465102 (arXiv:1107.4205)

\bibitem{Tarasov-Map2} Tarasov, V.E., Discrete map with memory from fractional differential equation of arbitrary positive order. \textit{Journal of Mathematical Physics}. \textbf{2009}. Vol.50. No.12. Article number: 122703. DOI: 10.1063/1.3272791 (arXiv:1107.4425)

\bibitem{Tarasov-Map3} Tarasov, V.E., Fractional Zaslavsky and Henon discrete maps. in \textit{Long-range Interactions, Stochasticity and Fractional Dynamics}. Edited by A.C.J. Luo, V. Afraimovich. New York: Springer and HEP, 2010. 275 pages. 
pp.1-26. DOI: 10.1007/978-3-642-12343-6\_1 (arXiv:1107.5148) 

\bibitem{TT-Logistic} Tarasova, V.V.; Tarasov, V.E., Logistic map with memory from economic model. \textit{Chaos, Solitons and Fractals}. \textbf{2017}. Vol.95. P.84-91. 
DOI: 10.1016/j.chaos.2016.12.012 (arXiv:1712.09092)

\bibitem{Entropy2021} Tarasov, V.E., 
Quantum maps with memory from generalized Lindblad equation.
\textit{Entropy}. \textbf{2021}. Vol.23. No.5. Article number: 544.
DOI: 10.3390/e23050544

\bibitem{Mathematics2021} Tarasov, V.E., 
Integral equations of non-integer orders and discrete maps with memory. 
\textit{Mathematics}. \textbf{2021}. Vol.9. No.11. Article number: 1177. 
DOI: 10.3390/math9111177 

\bibitem{MMAS2021} Tarasov, V.E., Predator-prey models with memory and kicks: Exact solution and discrete maps with memory.
\textit{Mathematical Methods in the Applied Sciences}. 
\textbf{2021}. Vol.44. DOI: 10.1002/mma.7510
Accepted for publication.

\bibitem{Mathematics2021-2} Tarasov, V.E., 
General fractional dynamics.
\textit{Mathematics}. \textbf{2021}. Vol.9. 
Accepted for publication.


\bibitem{TT-Edelman1} Tarasov, V.E.; Edelman, M., Fractional dissipative standard map. \textit{Chaos: An Interdisciplinary Journal of Nonlinear Science}. 
\textbf{2010}. Vol.20. No.2. Article number: 023127.
DOI: 10.1063/1.3443235

\bibitem{TT-Edelman2} Edelman, M.; Tarasov, V.E., Fractional standard map. \textit{Physics Letters A}. \textbf{2009}. 
Vol.374. No.2. P.279-285.
DOI: 10.1016/j.physleta.2009.11.008


\bibitem{Edelman1} Edelman, M., Fractional standard map: Riemann-Liouville vs. Caputo. \textit{Communications in Nonlinear Science and Numerical Simulation}. \textbf{2011}. 
Vol.16. No.12. P.4573-4580.
DOI: 10.1016/j.cnsns.2011.02.007

\bibitem{Edelman2} Edelman, M., Fractional maps and fractional attractors. Part I: alpha-families of maps. \textit{Discontinuity, Nonlinearity, and Complexity}. \textbf{2013}. 
Vol.1. No.4. P.305-324. DOI: 10.5890/DNC.2012.07.003 
(arXiv:1209.5681)

\bibitem{Edelman3} Edelman, M., Universal fractional map and cascade of bifurcations type attractors. \textit{Chaos: An Interdisciplinary Journal of Nonlinear Science}. 
\textbf{2013}. Vol.23. No.3. Article number: 033127.
DOI: 10.1063/1.4819165

\bibitem{Edelman5} Edelman, M., Caputo standard alpha-family of maps: fractional difference vs. fractional. \textit{Chaos: An Interdisciplinary Journal of Nonlinear Science}. 
\textbf{2014}. Vol.24. No.2. Article number: 023137. 
DOI: 10.1063/1.4885536 (arxiv:1406.4059)

\bibitem{Edelman7} Edelman, M., Fractional maps and fractional attractors. Part II: Fractional difference $\alpha$-families of maps. \textit{Discontinuity, Nonlinearity, and Complexity}. 
\textbf{2015}. Vol.4. P.391-402. 
DOI: 10.5890/DNC.2015.11.003 (arXiv:1404.4906)

\bibitem{Edelman-Handbook-2} Edelman, M., 
Maps with power-law memory: direct introduction and Eulerian numbers, fractional maps, and fractional difference maps.
In: {Handbook of Fractional Calculus with Applications. Volume 2. Fractional Differential Equations.} 
Edited by A. Kochubei, Yu. Luchko. 
Berlin, Boston: Walter de Gruyter GmbH, 2019. 519 pages. pp.47-63. ISBN: 978-3-11-057082-3
DOI: https://doi.org/10.1515/9783110571660-003

\bibitem{Edelman-Handbook-4} Edelman, M., 
Dynamics of nonlinear systems with power-law memory. In:
\textit{Handbook of Fractional Calculus with Applications. Volume 4. Application in Physics. Part A.} 
Edited by V.E. Tarasov. Berlin, Boston: Walter de Gruyter GmbH, 2019. 306 pages. pp.103-132. ISBN: 978-3-11-057088-5 
DOI: https://doi.org/10.1515/9783110571707-005 

\bibitem{Edelman2021} Edelman, M.
Cycles in asymptotically stable and chaotic fractional maps.
\textit{Nonlinear Dynamics}. \textbf{2021}. 
DOI: 10.1007/s11071-021-06379-2
13 pages. (arXiv:2010.12924)



\bibitem{CNSNS-2019-LAG} Tarasov, V.E.; Tarasova, S.S.,
Fractional and integer derivatives with continuously distributed lag.
\textit{Communications in Nonlinear Science and Numerical Simulation}. 
\textbf{2019}. Vol.70. P.125-169. DOI: 10.1016/j.cnsns.2018.10.014

\bibitem{PFDA-2019} Tarasov, V.E.; Tarasova, S.S.,
Probabilistic interpretation of Kober fractional integral of non-integer order.
\textit{Progress in Fractional Differentiation and Applications}.
\textbf{2019}. Vol.5. No.1. P.1-5. DOI: 10.18576/pfda/050101


\bibitem{GenFunc1} Lighthill, M.J., 
\textit{An Introduction to Fourier Analysis and Generalised Functions}. 
Cambridge: Cambridge University Press, 1958. 79 pages.
ISBN: 9781139171427 DOI: 10.1017/CBO9781139171427

\bibitem{GenFunc2} Gel'fand, I.M.; Shilov, G.E., 
\textit{Generalized Functions. Vol. I: Properties and Operations}. Boston: Academic Press, 1964. 
and American Mathematical Society, 2016.
423 pages. ISBN: 1-4704-2658-7 ISBN: 978-1-4704-2658-3


\bibitem{TS-1} 
Scher, H.; Shlesinger, M.F.; Bendler, J.T.,
Time-scale invariance in transport and relaxation. 
\textit{Physics Today}. 
\textbf{1991}. Vol.44. No.1. P.26-34. DOI: 10.1063/1.881289

\bibitem{TS-2} 
Lee, D.; Watkins, J.; Frame, D.; Given, G.; He, R.; Li, N.; Lu, B-N.; Sarkar, A.,
Time fractals and discrete scale invariance with trapped ions.
\textit{Physical Review A}. \textbf{2019}. Vol.100. No.1. Article number: 011403.
DOI: 10.1103/PhysRevA.100.011403 (arXiv:1901.01661)

\bibitem{TS-3}
Yulmetyev, R.M.; Mokshin, A.V.; Hanggi, P.; Shurygin, V.Yu., 
Time-scale invariance of relaxation processes of density 
fluctuation in slow neutron scattering in liquid cesium.
\textit{Physical Review E}. \textbf{2001}. Vol.64. No.5-2. Article number: 057101.
DOI: 10.1103/PhysRevE.64.057101

\bibitem{TS-4}
Stanley, H.E.; Amaral, L.A.N.; Gopikrishnan, P.; Plerou, V.; Salinger, M.A.,
Scale invariance and universality in economic phenomena.
\textit{Journal of Physics: Condensed Matter}. 
\textbf{2002}. Vol.14. No.9. Article number: 2121.

\bibitem{TS-5}
Feigenbaum, J.A.; Freund, P.G.O.,
Discrete scale invariance in stock markets before crashes.
\textit{International Journal of Modern Physics B}. 
\textbf{1996}. Vol.10. No.27. P.3737-3745.
DOI: 10.1142/S021797929600204X

\bibitem{TS-6}
Proekt, A.; Banavar, J.R.; Maritan, A.; Pfaff, D.W.,
Scale invariance in the dynamics of spontaneous behavior.
\textit{Proceedings of the National Academy of Sciences (PNAS)}. 
\textbf{2012}. Vol.109. No.26. P.10564-10569. 
DOI: 10.1073/pnas.1206894109 

\bibitem{TS-7}
Buhusi, C.V.; Oprisan, S.A.,
Time-scale invariance as an emergent property in a perceptron with realistic, noisy neurons.
\textit{Behavioural Processes}. \textbf{2013}. Vol.95. P.60-70. 
DOIi: 10.1016/j.beproc.2013.02.015

\bibitem{TS-8}
Khaluf, Y.; Ferrante, E.; Simoens, P.; Huepe, C.,
Scale invariance in natural and artificial collective systems: a review.
\textit{Journal of the Royal Society Interface}. 
\textbf{2017}. Vol.14. No.136. Article number: 20170662.
DOI: 10.1098/rsif.2017.0662


\end{thebibliography}
\end{document}